\documentclass[12pt]{article}
\pdfoutput=1

\usepackage[usenames,dvipsnames,svgnames,table]{xcolor} 
\usepackage[obeyspaces,hyphens,spaces]{url}
\usepackage{jcapmod}
\usepackage{verbatim}

\usepackage{booktabs}
\usepackage[english]{babel}
\usepackage{amsmath, amssymb, amsbsy, amstext, amsthm, simplewick}
\usepackage{hyperref}
\usepackage{graphicx}
\usepackage{amsfonts}
\usepackage{upgreek}
\usepackage{framed}
\usepackage{tensor}
\usepackage{pifont}
\usepackage{latexsym, mathrsfs}
\usepackage{array}
\usepackage{hyperref}
\usepackage{xspace}
\usepackage{longtable}
\usepackage{multirow}
\usepackage{cases}
\usepackage{empheq}

\usepackage{bm}
\usepackage{bbm}

\usepackage{afterpage}
\usepackage{setspace}

\usepackage{braket}

\usepackage[load-configurations=astronomy, range-units=brackets, range-phrase=-, per-mode=reciprocal, mode=math]{siunitx}
\usepackage{threeparttable}
\usepackage{subfig}

\usepackage{ragged2e}

\usepackage{hhline}
\usepackage{array}
\newcolumntype{P}[1]{>{\centering\arraybackslash}p{#1}}
\usepackage{booktabs}
\usepackage{tikz}
\usetikzlibrary{calc,shadings,patterns,tikzmark,fadings}

\definecolor{Blue}{rgb}{0.25, 0.41, 0.88}
\definecolor{Red}{rgb}{0.92,0.,0.}
\definecolor{darkorange}{rgb}{1.0,0.549,0.}
\definecolor{cobalt}{RGB}{44, 98, 120}
\definecolor{Mathematica1}{rgb}{0.368417, 0.506779, 0.709798}
\definecolor{Mathematica2}{rgb}{0.880722, 0.611041, 0.142051}
\definecolor{Mathematica3}{rgb}{0.560181, 0.691569, 0.194885}
\definecolor{Mathematica4}{rgb}{0.922526, 0.385626, 0.209179}
\definecolor{Mathematica5}{rgb}{0.528488, 0.470624, 0.701351}
\definecolor{Mathematica6}{rgb}{0.772079, 0.431554, 0.102387}
\definecolor{Mathematica7}{rgb}{0.363898, 0.618501, 0.782349}
\definecolor{Mathematica8}{rgb}{1, 0.75, 0}
\definecolor{Mathematica9}{rgb}{0.647624, 0.37816, 0.614037}
\definecolor{plotBlue}{RGB}{94, 130, 181}
\definecolor{plotRed}{RGB}{233, 85, 54}
\definecolor{plotGreen}{RGB}{142, 176, 50}
\definecolor{plotPurple}{RGB}{135, 120, 178}

\newcolumntype{C}[1]{>{\centering\let\newline\\\arraybackslash\hspace{0pt}}m{#1}}

\def\x{{\bf x}}

\makeatletter
\newlength{\apb@width}
\newcommand{\autoparbox}[2][c]{\settowidth{\apb@width}{#2}\parbox[#1]{\apb@width}{#2}}

\makeatother

\makeatletter
\newsavebox\myboxA
\newsavebox\myboxB
\newlength\mylenA

\newcommand*\xoverline[2][0.75]{
    \sbox{\myboxA}{$\m@th#2$}%
    \setbox\myboxB\null
    \ht\myboxB=\ht\myboxA%
    \dp\myboxB=\dp\myboxA%
    \wd\myboxB=#1\wd\myboxA
    \sbox\myboxB{$\m@th\overline{\copy\myboxB}$}
    \setlength\mylenA{\the\wd\myboxA}
    \addtolength\mylenA{-\the\wd\myboxB}%
    \ifdim\wd\myboxB<\wd\myboxA%
       \rlap{\hskip 0.5\mylenA\usebox\myboxB}{\usebox\myboxA}%
    \else
        \hskip -0.5\mylenA\rlap{\usebox\myboxA}{\hskip 0.5\mylenA\usebox\myboxB}%
    \fi}
\makeatother

\numberwithin{equation}{section}

\def\beq{\begin{equation}}
\def\eeq{\end{equation}}

\def\bea{\begin{eqnarray}}
\def\eea{\end{eqnarray}}

 \def\be{\begin{equation}}
 \def\ee{\end{equation}}
 \def\bes{\begin{eqnarray}}
 \def\ees{\end{eqnarray}}

\def\beq{\begin{equation}}
\def\eeq{\end{equation}}
\def\bea{\begin{eqnarray}}
\def\eea{\end{eqnarray}}

\DeclareRobustCommand{\SkipTocEntry}[4]{}

\usepackage{colortbl}
\definecolor{blue2}{cmyk}{1, 0.1, 0.1, 0.1}

\definecolor{pyBlue}{RGB}{31, 119, 180}
\definecolor{pyRed}{RGB}{214, 39, 40}
\definecolor{pyGreen}{RGB}{44, 160, 44}
\definecolor{pyBlue2}{RGB}{0, 111, 237}
\definecolor{pyRed2}{RGB}{224, 52, 36}

\usepackage{tikz,xcolor,hyperref}

\usepackage{slashed}

\definecolor{lime}{HTML}{A6CE39}
\DeclareRobustCommand{\orcidicon}{
	\begin{tikzpicture}
	\draw[lime, fill=lime] (0,0) 
	circle [radius=0.2] 
	node[white] {{\fontfamily{qag}\selectfont \tiny ID}};
	\draw[white, fill=white] (-0.0625,0.095) 
	circle [radius=0.007];
	\end{tikzpicture}
	\hspace{-2mm}
}

\foreach \x in {A, ..., Z}{\expandafter\xdef\csname orcid\x\endcsname{\noexpand\href{https://orcid.org/\csname orcidauthor\x\endcsname}
			{\noexpand\orcidicon}}
}

\begin{document}

\pagenumbering{roman}
\begin{titlepage}
\baselineskip=14.5pt \thispagestyle{empty}

\bigskip\

\vspace{0cm}
\begin{center}
{\fontsize{40}{40}\selectfont  \bfseries \textcolor{Sepia}{\vspace{0.3cm}${\cal N}$otes ${\cal O}$n}} \\ {\fontsize{40}{40}\selectfont  \bfseries \textcolor{Sepia}{\vspace{0.3cm} de-${\cal S}$itter}}  {\fontsize{40}{40}\selectfont  \bfseries \textcolor{Sepia}{${\cal M}$ellin ${\cal B}$arnes  ${\cal A}$mplitudes}}
\end{center}
\vspace{0.1cm}
\begin{center}
	
	{\fontsize{14}{18}\selectfont Sayantan Choudhury\orcidA{}${}^{\textcolor{Sepia}{1}}$\footnote{{\sffamily \textit{ Corresponding author, E-mail}} : {\ttfamily sayantan\_ccsp@sgtuniversity.org,  sayanphysicsisi@gmail.com}}}${{}^{,}}$
	\footnote{{\sffamily \textit{ NOTE: This project is the part of the non-profit virtual international research consortium ``Quantum Aspects of Space-Time \& Matter" (QASTM)} }. }${{}^{,}}$.~

\end{center}

\begin{center}
\vskip8pt
	\textit{${}^{1}$Centre For Cosmology and Science Popularization (CCSP),
SGT University, Gurugram, Delhi-NCR, Haryana- 122505, India.}\\
	


\end{center}

\vspace{0.09cm}
\hrule \vspace{0.09cm}
\begin{center}
\noindent {\bf Abstract}\\
\end{center}
In this paper, we create a Mellin space method for boundary correlation functions in de Sitter (dS) and anti-de Sitter (AdS) spaces. We demonstrate that the analytic continuation between AdS${}_{d+1}$ and dS${}_{d+1}$ is encoded in a set of simple relative phases using the Mellin-Barnes representation of correlators. It helps us to determine the scalar three-point and four-point functions and their corresponding Mellin-Barnes amplitudes in dS${}_{d+1}$ space using the known results from AdS${}_{d+1}$ space. The Mellin-Barnes representation reveals the analytic structure of boundary correlation functions over all $d$ and scaling dimensions. In the present discussion, the {\it split representation} have been used as an instrumental technique in particularly the evaluation of bulk
Witten diagrams and is suitable to obtain the {\it Conformal
Partial Wave decomposition} of tree-level exchange in the bulk Witten diagrams. The equivalent adjustment to the cosmological three-point and four-point function of generic external scalars may be further extracted from these results, assuming the weak breakdown of the de Sitter isometries. These findings offer a step towards a more methodical comprehension of de Sitter observables utilising Mellin space techniques at the tree level and beyond.

\vskip10pt
\hrule
\vskip10pt

\text{Keywords:~~Scattering Amplitudes,~QFT of de Sitter space,~ Cosmology.}

\end{titlepage}

\thispagestyle{empty}
\setcounter{page}{2}
\begin{spacing}{1.03}
\tableofcontents
\end{spacing}

\clearpage
\pagenumbering{arabic}
\setcounter{page}{1}

\clearpage

\section{Introduction}

 A formalism that makes the dynamics and symmetries obvious and straightforward is frequently necessary for advancements in physics. For instance, recent advances in S-Matrix theory have been made possible by the use of twistor space, on-shell superspace, and the spinor-helicity formalism. We shall contend that the most natural framework for CFT \cite{Maldacena:1997re,Gubser:1998bc,Heemskerk:2009pn,El-Showk:2011yvt} correlation functions is the Mellin representation \cite{Mack:2009mi,Mack:2009gy,Penedones:2010ue,Paulos:2011ie,Fitzpatrick:2011ia,Fitzpatrick:2011dm,Fitzpatrick:2011hu,Costa:2012cb,Goncalves:2014rfa,Sleight:2019hfp,Sleight:2019mgd,Sleight:2020obc}. The advantages of using  Mellin space for correlation functions and scattering amplitudes in flat spacetime are fundamentally analogous to the important but more pedestrian transition from position to momentum space. This relationship has been made possible by the Mellin space representation of conformal correlators and Harmonic Analysis for the Euclidean Conformal Group \cite{Costa:2012cb,Mack:1974sa,Dobrev:1975ru,Dobrev:1977qv,Caron-Huot:2017vep}, which express bulk physics in a manner that has important parallels with the flat-space scattering amplitudes.

 Witten diagrams give us the ability to calculate correlation functions of strongly coupled conformal field theories with a gravity dual, but despite tremendous advancements, these calculations are generally very difficult to carry out. Currently, the state of the art is the computation of four point functions involving various types of exchanged fields in type IIB supergravity and a stress-tensor three-point function, which is an especially heroic effort because of the complex tensor structures. Coordinate space is typically used for these computations. One obvious question is whether simplifications may result from modifying the foundation. The initial guess is momentum space, however this doesn't result in any significant simplifications. This might be because such a transformation just considers the border of AdS space, not its symmetries. It turns out that there is a better suitable basis: the Mellin transform should be used in place of the Fourier transform. 

 On the other hand, we are still in the preliminary stages of comprehending border correlators in de Sitter space \cite{Guth:1980zm,Linde:1981mu,Albrecht:1982wi,Starobinsky:1982ee}. These are spatial correlations at late periods that encode the traces of previous scattering events, as opposed to scattering amplitudes. Correlators in the dual Euclidean CFT are therefore not required to meet the Osterwalder-Schrader axioms, including reflection positivity. The principles that the associated late-time correlators must follow, particularly how they convey continuous bulk time evolution, are now beyond our comprehension. The propagators' dependence on conformal time follows a simple power-law at the Mellin-Barnes representation, making bulk integrals computationally straightforward. This feature generates analytic formulas for border correlators with any number of legs. The Mellin-Barnes representation of boundary correlators reveals their analytic structure, including momenta, boundary dimension (d), and field scaling dimensions. Methods from the Mellin-Barnes literature can be used to calculate asymptotic expansions of correlations. The Mellin-Barnes representation offers a useful framework for investigating the fundamental principles of late-time de Sitter correlators, which may be used to bootstrap observables without relying on bulk time evolution. Conformal symmetry determines the placement of the poles in the Mellin-Barnes integrand, whereas suitable boundary conditions can fix the zeros at singularities.

\begin{figure*}[htb]
    	\centering
    	{
    		\includegraphics[width=15.5cm,height=9.5cm] {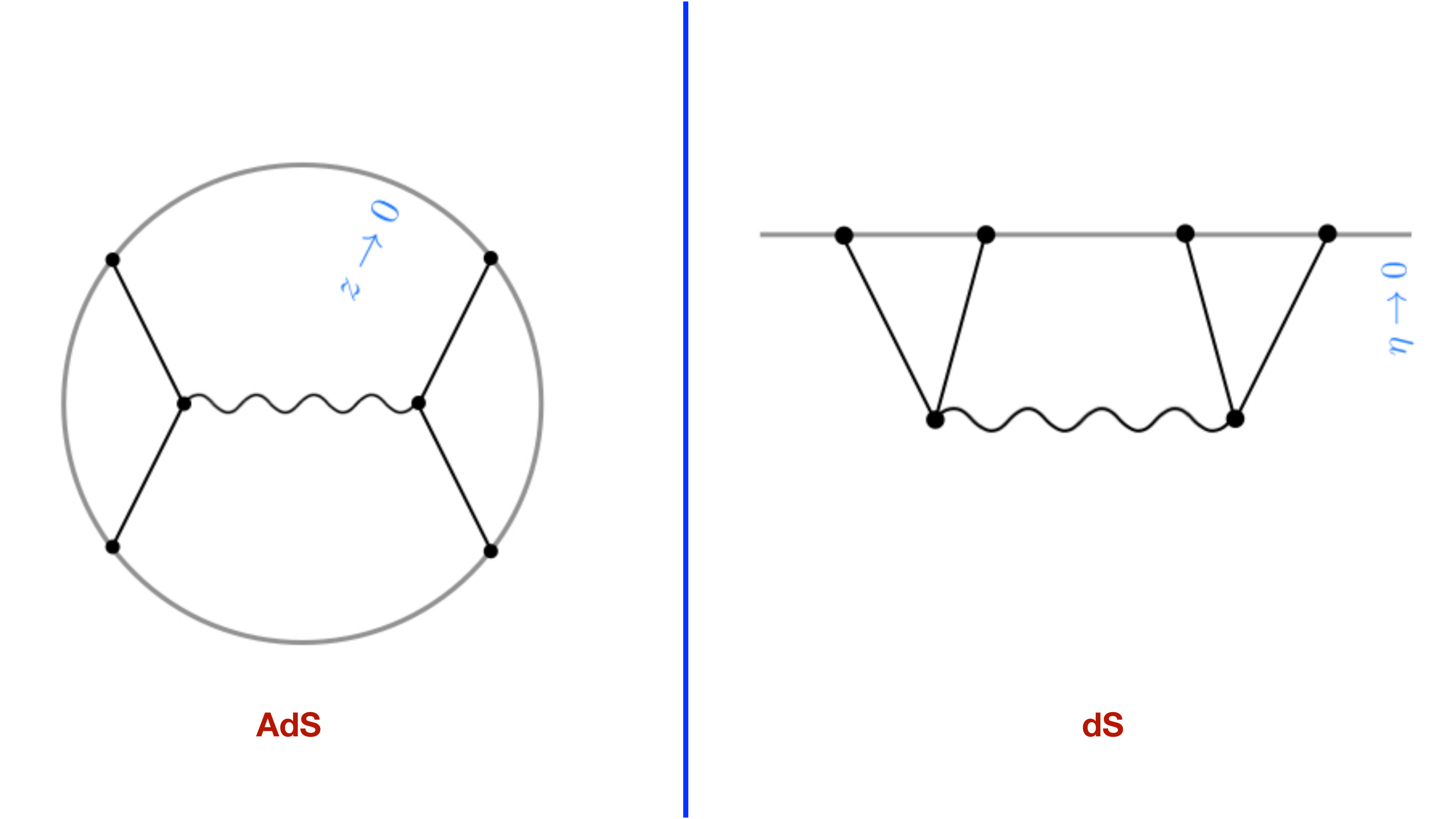}
    	}
    	\caption[Optional caption for list of figures]{Representative diagram showing comparison between Anti de Sitter (AdS) and de Sitter (dS) scattering amplitudes at the four-point level.} 
    	\label{Overall}
    \end{figure*}

 The work examined CFT correlation functions calculated in the AdS/CFT and dS/CFT contexts using the Mellin formalism, with encouraging outcomes. In contrast to the complex D-functions that arise in coordinate space, contact interactions have Mellin amplitudes that are simple polynomials. In the case of sparsely connected scalars, even the feared stress-tensor exchange diagram simplifies to a straightforward rational function. The explicit gamma functions in the Mellin representation capture double-trace operators corresponding to the fusion of external legs, while single-trace operators and their descendants corresponding to internal lines or bulk-to-bulk propagators appear as simple poles of the Mellin Barnes amplitude. These basic analytic properties of Mellin amplitudes also reveal which operators are propagating throughout a given Witten diagram in AdS space and a Witten-like diagram in dS space. We will use the Mellin framework to compute tree-level correlation functions of generic scalars on (d+1)-dimensional de Sitter space. This includes n-point contact diagrams and four-point exchange diagrams. From the observational point of view, computations performed particularly in the dS/CFT perspective is extremely relevant in the context of the study of primordial cosmological correlations. Using the Mellin-Barnes representation of quasi-dS correlation functions with $d=3$, one can study various unexplored features of small and large primordial fluctuations, which are directly related to the study of the inflationary paradigm \cite{Choudhury:2025hnu,Choudhury:2025vso,Choudhury:2011sq,Choudhury:2012yh,Choudhury:2013zna,Choudhury:2013jya,Choudhury:2013iaa,Choudhury:2014sxa,Choudhury:2014uxa,Choudhury:2014kma,Choudhury:2014sua,Choudhury:2015pqa,Choudhury:2015hvr,Choudhury:2017cos} and primordial black hole formations \cite{Choudhury:2025kxg,Choudhury:2024kjj,Choudhury:2024aji,Choudhury:2024dzw,Choudhury:2024dei,Choudhury:2024jlz,Choudhury:2024ybk,Choudhury:2024one,Choudhury:2023fjs,Choudhury:2023fwk,Choudhury:2023hfm,Choudhury:2023kdb,Choudhury:2023hvf,Choudhury:2023rks,Choudhury:2023jlt,Choudhury:2023vuj,Choudhury:2013woa,Choudhury:2011jt}. Though in this paper we have not directly computed such higher-point cosmological correlation functions, our results obtained for the dS/CFT correlation functions can be further extended to explore various unaddressed issues in the present context of discussion. There are another couple of directions along which one can further extend the results obtained for dS/CFT correlators, eg. study of cosmological collider signals \cite{Chen:2009zp,Maldacena:2011nz,Baumann:2011nk,Assassi:2012zq,Chen:2012ge,Noumi:2012vr,Assassi:2013gxa,Arkani-Hamed:2015bza,Lee:2016vti,An:2017hlx,Kumar:2017ecc,Baumann:2017jvh,Goon:2018fyu}  in terms of non-Gaussian cosmological correlation functions and, last but not the least, the non-perturbative treatment of bootstrapping cosmological correlators \cite{Arkani-Hamed:2018kmz}.
In figure \ref{Overall}, we have shown a representative diagram through which one can visualize the omparison between Anti de Sitter (AdS) and de Sitter (dS) scattering amplitudes at the four-point level.

 The organization of this paper is as follows. In section \ref{sec:prelim}, we start our discussion with the preliminaries, where we explicitly mention all the required and relevant notations and conventions, which we are going to use very frequently in the rest of the papers. Further, in section \ref{sec:embed}, we give a very short overview on the embedding formalism, which is basically the building block of the rest of the computations performed in this paper. Next, in section \ref{sec:geo}, we review the underlying geometry of the $d+1$ dimensional de Sitter (dS) and Anti de Sitter (AdS) space-time, which is necessarily required for the desired correlators and the related amplitudes. Then, in section \ref{sec:gn}, we give a general overview of the Mellin-Barnes transformation and its applications in the context of computing correlation functions and related amplitudes in AdS and dS space-time. Further, in the subsections \ref{sec:2ds} and \ref{sec:2ads}, we explicitly compute the propagators and the associated two-point functions in $d+1$-dimensional dS and AdS space-time. Next, in the subsections \ref{sec:3ads} and \ref{sec:3ds},  we explicitly compute the three-point function and the associated amplitudes using the Mellin-Barnes representation in $d+1$-dimensional dS and AdS space-time. Further, in the subsections \ref{sec:4ads} and \ref{sec:4ds},  we explicitly compute the four-point function and the associated amplitudes using the Mellin-Barnes representation in $d+1$-dimensional dS and AdS space-time \footnote{One can, in principle, compute the higher point correlators and the associated Mellin-Barnes amplitudes in AdS and dS space. However, due to having the tight constraints from cosmological observations, finding out the expressions for the higher-point cosmological correlators, more than the four-point, is not physically relevant. Since the results obtained for dS Mellin Barnes amplitudes are cosmologically relevant, in the present context of discussion, we have restricted our computation to four-point function.}. Finally, in section \ref{sec:conclu}, we conclude with future prospects.
 
\section{Preliminaries: Notation and convention}
\label{sec:prelim}
The notations and conventions used in this paper are appended below point-wise:
\begin{enumerate}
\item We primarily work in $(d + 1)$-dimensional De Sitter space-time with the metric signature
$(- + +... +)$. 

\item In this context, the Greek letters denote $(d+1)$-dimensional De Sitter space-time indices, $\mu = 0, 1, ..., d$, lower-case
Latin letters denote $d$-dimensional spatial part of the $(d+1)$-dimensional De Sitter space-time, $i = 1, ..., d$, while $(d+2)$ dimensional ambient Minkowski space-time indices are denoted
by $M,N = 0, 1, ..., d + 1$.

\item Bulk scalar fields of scaling dimension, \bea \Delta_{\pm}=\frac{d}{2}\pm i\nu,\eea
are represented by $\phi^{\pm}_{(\nu)}$ and the arbitrary spin fields with spin-$s$ are denoted by $\phi^{\pm}_{s,(\nu)}$.
` .
\item We will also use the natural units for which we take $\hbar=1$ and $c=1$.
\end{enumerate}
\section{Embedding formalism}
\label{sec:embed}
In this section, we will discuss some preliminaries. (A)dS correlators can be written in embedding space formalism in which  (A)dS$_{d+1}$ space is seen as a curved surface embedded in flat Minkowski space in one higher dimension ($\mathbb{M}_{d+2}$).\footnote{It is a curious historical fact that Dirac thought about this idea in 1930s.}
\bea
\label{eqn1}
ds^2 &=&\eta_{AB} dX^A dX^B \nonumber\\
&=& - (dX^0)^2 + \sum^{d+2}_{i=1}(dX^i)^2 \nonumber\\
&=&-dX^+ dX^- +\delta_{mn}dX^m dX^n= l^2=\frac{1}{H^2}.
\eea
where, we define:
\be \eta_{AB} = \text{diag} (-1,1 \dots 1,1)~\forall~A,B=0,\cdots,d+1.\ee The constant $l=H^{-1}$ is the de-Sitter radius. With analytic continuation  \cite{Maldacena:2002vr,Ghosh:2014kba,Anninos:2014lwa}, de-Sitter embeddings can be obtained from a sphere:
\be
\label{eqn2}
 \sum^{d+2}_{i=1}(dX^i)^2 = l^2=\frac{1}{H^2}, 
\ee
by analytic continuation,
\be
dX^{d+2} \mapsto \pm i X^0
\ee
or from Euclidean AdS,
\begin{equation}
X^M \mapsto \pm i X^M
\end{equation}
In this formalism, it is convenient to think of the conformal boundary is identified with light rays $P^A$ (with
$P^2=0$, and $P\sim \lambda P$). Then a correlation function of the dual CFT of weight $\Delta$ scales as:
\be  \mathcal{F}_\Delta (\lambda P) = \lambda^{-\Delta} \mathcal{F}_\Delta (P).\ee
 In the light cone coordinate, one can write:
\bea
\label{eqn2x}
X^A:&=&(X^+,X^-,X^{\mu})=\frac{1}{x_0}(1,x^2_0+x^2,x^{\mu}),\\
P^A:&=&(P^+,P^-,P^\mu)=(1,y^2,y^\mu),
\eea
where $\mu=0,1,\cdots,d-1$.  Here, $x^{\mu}$ and $y^{\mu}$ are the $d$-dimensional vectors whose lengths are defined as:
\be x^2=x_{\mu}x^{\mu},~~~~~y^2=y_{\mu}y^{\mu}.~~~\ee
In the remainder of the paper we will frequently use the following notation,
We will be interested in $n$-point correlation functions of the form $\mathcal{F} (P_1, P_2, \dots , P_n)$, and frequently use the notation
\bea
\label{eqn3}
&&-2 P\cdot X=\frac{1}{x_0}\left(x^2_0+(x-y)^2\right),\\
&&-2P_i \cdot P_j \ =P_{ij}=(y_i-y_j)^2.
\eea
We will use $X^A$, $Y^A$, etc., for points in the bulk, and $P^A$, $Q^A$, etc., for points on the boundary of (A)dS space.

\section{On the geometry of de Sitter (dS) and Anti de Sitter (AdS) space}
\label{sec:geo}
One can easily think of the de Sitter space as the embedding:
\begin{equation}
    -(X^0)^2+ \sum_{i=1}^{d+1} (X^i)^2=\mathcal{\ell}^2=\frac{1}{H^2}
\end{equation}
into a $(d+2)$-dimensional Minkowski space-time represented by the metric:
\be ds^2_{d+2}=\eta_{MN}dX^M dX^N~~~~{\rm where}~~~~\eta_{MN}=\text{diag} (-,+\dots,+,+)~~\forall M, N=0,...,d+1.\ee  Also, $\ell=-1/H$ is the radius of the de Sitter space and hence represented by a constant, which is the inverse of the Hubble constant $H$.\footnote{Note that the de Sitter embedding can be obtained from the sphere $S^{d+1}$: \be \displaystyle\sum_{i=1}^{d+1} X_i^2=\mathcal{\ell}^2+ X_{d+2}^2=\ell^2=\frac{1}{H^2} \ee by analytically continuing
 $X^{d+2}\to \pm i X^0$.} We will work in the Poincare patch (flat slicing coordinates), and the line element can be written as:
\begin{equation}
ds^2_{d+1}=a^2(\tau)\left(-d\tau^2+d\vec{x}^2\right)~~~~{\rm with}~~a(\tau)=\frac{l}{\tau}=-\frac{1}{H\tau}.
\end{equation}
where $\tau$ is the conformal time coordinate which can be expressed in terms of the usual time coordinate as: $d \tau=dt/a(t)$~\footnote{In terms of the usual time coordinate the de Sitter metric in the Poincare patch (flat slicing coordinates) can be expressed as:
\be ds^2_{d+1}=-dt^2+a^2(t)d\vec{x}^2~~~~{\rm with}~~a(t)=\exp(Ht)\ee} and the above is obtained from a $(d+2)$-dimensional Minkowski space-time by parametrizing in the following way:
\begin{equation}
X^M=\frac{\ell}{\tau}\left[\frac{\ell^2-(\tau^2- \vec{x}^2)}{2\ell},~\vec{x},~\frac{\ell^2+(\tau^2- \vec{x}^2)}{2\ell}\right]
\end{equation}
 where $\tau$ is the conformal time and the $\vec{x}$ represents the 
$d$ dimensional spatial slices including the late time conformal boundary which is at $\tau=0$. For comparison, we can also think about solving the problem in Euclidean AdS (EAdS) signature~\footnote{One can transform a $(d+2)$ dimension Minkowski embedding space-time to a $(d+2)$ dimensional EAdS embedding space-time by considering the Wick rotation in the conformal time coordinate, $\tau\rightarrow \pm iz$.} where one can translate the above mentioned embedding coordinate as:
\begin{equation}
Y^M=\frac{\ell}{z} \left[ \frac{\ell^2 + (z^2 + \vec{x}^2)}{2\ell},~\vec{x},~\frac{\ell^2- (z^2+ \vec{x})}{2\ell} \right].
\end{equation}
Here the radial bulk coordinates $\vec{x}$ parametrizes the conformal boundary of Anti de Sitter (AdS) space at $z=0$. The conformal boundary is identified by the following constraint condition:
\be P^2 =0~~ {\rm with}~~ P \sim \alpha P~~{\rm  where}~~ \alpha \neq 0.\ee 
    \begin{figure*}[htb]
    	\centering
    	{
    		\includegraphics[width=15.5cm,height=9.5cm] {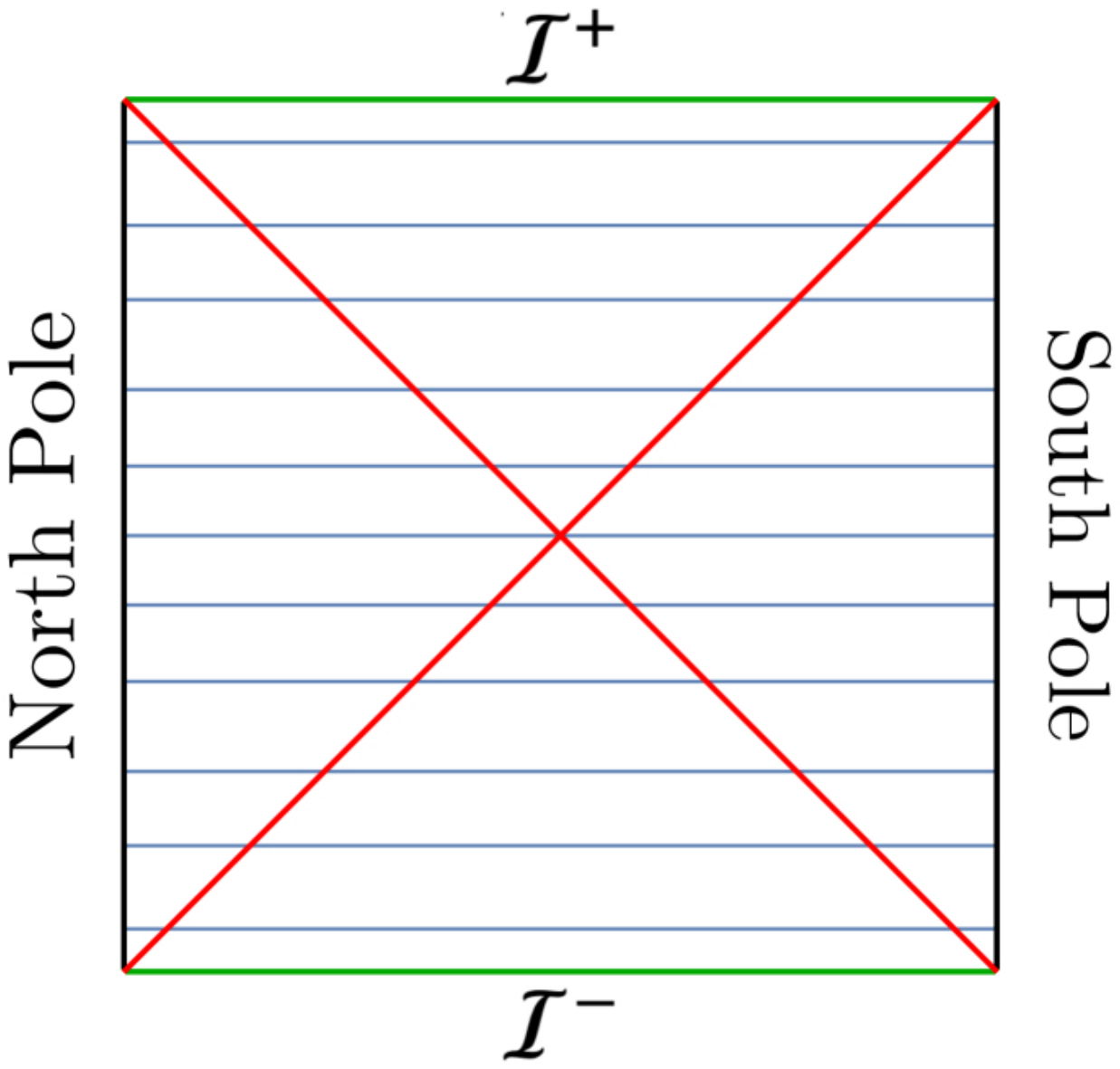}
    	}
    	\caption[Optional caption for list of figures]{Representative picture showing the Penrose diagram of de Sitter space.} 
    	\label{dS}
    \end{figure*}
In this context, the ambient boundary point can be written as:
\begin{equation}
Y^M \rightarrow P^M = \frac{1}{2} (1 + x^2,~ 2 \vec{x},~ 1-x^2)
\end{equation}
For convenience, we can write down, 
\begin{equation}
 P \cdot X = \frac{-x^2+2 x y+\left(\eta ^2-1\right) y^2}{2 \eta }
 \end{equation}
 From this ingredient, we can compute the following bulk to boundary propagator in the Witten diagrams and we will discuss this later:
 \begin{align}
 K_{\Delta,0}(z,\vec{y}_1;\vec{y}_2)=C_{\Delta ,0} \left(\frac{z_1}{\left(y_1-y_2\right){}^2+z^2}\right)^{\Delta }\end{align}
Similarly,
\begin{equation}
X_1 \cdot X_2 =  -(x-y)^2
\end{equation}

In this discussion, the two point functions are the functions of the geodesic distance $D$, which is represented by:
\be \cos \left( \frac{D}{\ell}\right)=2\sigma-1,\ee
 where $\sigma$ represents the chordal distance. From this we can write down $\sigma$ for both Anti de Sitter (AdS) and de Sitter (dS) space which is given by:
   \be \sigma_{AdS} = \frac{1+ Y_1 \cdot Y_2}{2}=\frac{\left(z_1-z_2\right){}^2-x_{12}^2}{4 z_1 z_2},\ee
   \be
\sigma_{dS} = \frac{1+ X_1 \cdot X_2}{2}=-\frac{\left(\tau _1+\tau _2\right){}^2+x_{12}^2}{4 \tau _1 \tau _2}.\ee
where we define $x_{12} = |\vec{x}_1 - \vec{x}_2|$, as the distance between two $d$ dimensional vectors located at the spatially flat slices. 

Further, the analytic continuation which can can be written as:
 \bea z_1= \tau _1~ \exp\left(\frac{1}{2} (\pm i) \pi \right)~~~\Longrightarrow~~~Y_1=\mp i~X_1,\\
 z_2=\tau_2 ~\exp\left( \frac{1}{2} (\mp i) \pi\right)~~~\Longrightarrow~~~Y_2=\pm i~X_2.\eea 
 For this reason we will consider the Wightman two point functions in dS$_{d+1}$ as an analytic continuation of EAdS$_{d+1}$. Also, it is important to note that, in figure \ref{dS}, we have explicitly shown the Penrose diagram of de Sitter space, which is helpful to understand the underlying geometry and the scattering process in the corresponding dS patch of the Penorse diagram.

\section{General notes on Mellin Barnes transformation and its application to amplitudes} 
\label{sec:gn}
In the context of mathematics, {\it Mellin Barnes transformation} is treated as an integral transformation which incorporates the multiplicative version of the two sided {\it Laplace transformation}. The {\it Mellin Barnes transformation} of a function $f$ with a single variable $t$ is defined on the positive real axis, $f:\mathbb{R}_{+}\rightarrow \mathbb{R}$, is defined as:
\be g(s):=\int^{\infty}_{0}dt~t^{s-1}~f(t)~~~{\rm for~ a~strip}~{\bf S}:=\left\{s\in \mathbb{C}|a<{\cal R}(s)<b,~~b>a>0\right\}.\ee
Also, the {\it inverse Mellin Barnes transformation} of the function $f(t)$ with the sigle variable $t$ is defined by the following integral in the complex plane as:
\be f(t):=\frac{1}{2\pi i}\int^{c+i\infty}_{c-i\infty}ds~g(s)~t^{-s}~~~~~\forall~~b>c>a>0,\ee
where the function $f(t)$ is piecewise continuous on the positive real axis, provided for the {\it staircase function} one need to consider the two-sided limiting values at the discontinuous jump points.

One can further generalize this integral representation of {\it Mellin Barnes transformation} for $N$ number of variables, which is given by:
\be g(s_1,s_2,\cdots,s_N):=\int^{\infty}_{0}\int^{\infty}_{0}\cdots\int^{\infty}_{0}dt_1dt_2\cdots dt_N~t^{s_1-1}_1 t^{s_2-1}_2\cdots t^{s_N-1}_N~f(t_1,t_2,\cdots,t_N),\ee
and similarly the {\it inverse Mellin Barnes transformation} for the $N$ number of variables can be generalized by the following expression:
\bea &&f(t_1,t_2,\cdots,t_N):=\left(\frac{1}{2\pi i}\right)^N \int^{c_1+i\infty}_{c_1-i\infty}\int^{c_2+i\infty}_{c_2-i\infty}\cdots\int^{c_N+i\infty}_{c_N-i\infty}ds_1ds_2\cdots ds_N~g(s_1,s_2,\cdots,s_N)\nonumber\\
&&~~~~~~~~~~~~~~~~~~~~~~~~~~~~~~~~~~~~~~~~~~~~t^{-s_1}_1 t^{-s_2}_2\cdots t^{-s_N}_N~~~~~\forall~~c_i>0~~{\rm with}~i=1,2,\cdots,N,~~~~~~~~~~\eea
In the context of Conformal Field Theory (CFT) the {\it Mellin Barnes representation} for Euclidean correlator of primary scalar operators, $  \mathcal{O}_i(x_i),\forall i=1,2,\cdots,N$ having conformal dimension, $\Delta_{i}\forall i=1,2,\cdots,N$, can be written in terms of the {\it Mellin Barnes amplitude} by the following expression:
\bea
\mathcal{A}(x_1,x_2,\cdots,N):&=&\langle \mathcal{O}_1(x_1) \dots \mathcal{O}_N(x_N) \rangle\nonumber\\
&=& \frac{{\cal N}_{\rm norm}}{ (2 \pi i)^{\frac{N (N-3)}{2}}} \int d \delta_{ij}~ M(\delta_{ij})~ \prod_{i<j}^N \Gamma(\delta_{ij}) \left(x_{ij}^2 \right)^{-\delta_{ij}}
\eea
where the integration contour runs parallel to the imaginary axis with $\delta_{ij} >0$. In addition, these are constrained by the following condition:
\begin{equation}
\delta_{ij}=\delta_{ji},~~~\sum_{i,j,i \neq j}^N \delta_{ij} = \Delta_i,~~~~\delta_{ii}=-\Delta_{i}~\forall i=1,2,\cdots,N,~~ \sum_{i,j,i=j}^N \delta_{ij} =0,
\end{equation}
which may be solved by introducing a set of $d$-dimensional vectors $k_i$ which are characterized by:
\be \delta_{ij}:=k_{i}~{\bf{.}}~k_{j},~ ~k^2_i=k_{i}~{\bf{.}}~k_{i}=-\Delta_{i}~\forall i=1,2,\cdots,N,~~~~~\sum^{N}_{i=1} k_{i}=0.\ee
To make the computation simpler it is also useful to introduce {\it Mandelstam invariants} for $N$ point amplitude, which in the present context defined as:
\be s_{i_1 i_2 \cdots i_N}=-\left(\sum^{N}_{p=1}k_{i_p}\right)^2=\sum^{N}_{p=1}\Delta_{i_p}-2\sum^{N}_{k,l,~i_k<i_l}\Delta_{i_k j_l}\ee
Here the integrand is conformally covariant whose scaling dimension is $\Delta_i$ at point $x_i$.
One has to do $\frac{n (n-3)}{2}$ independent integration variables which is the same number of independent conformal invariant cross ratios for $N$ point correlators and hence represents number of independent $N$ particle scattering amplitude. In this discussion, ${\cal N}_{\rm norm}$ represents the normalization factor which we will fix during the computation in the context of de Sitter space. Also, $M(\delta_{ij})\forall i,j=1,2,\cdots,N ~(i\neq j)$ represents the {\it Mellin Barnes amplitude}, which is the prime object  of interest in this paper. We will show explicit examples of different {\it Mellin Barnes amplitudes} in the rest part of this paper. {\it Mellin Barnes amplitudes} have very simple analytic structures. In this paper, the {\it Mellin Barnes formalism} is used as a mathematical trick to study CFT correlation functions
computed in the context of dS/CFT which can be obtained further by performing analytical continuation from EAdS/CFT results very easily. 

\section{Two point function }
\subsection{From the scalar field in dS space}
\label{sec:2ds}
We will consider the scalar field $\Phi$ of mass $m$ represented by the following action:
\bea S&=&\int d^{d+1}x~\sqrt{-g_{(d+1)}}~\left[-\frac{1}{2}(\partial\Phi)^2+\frac{m^2}{2}\Phi^2\right]\nonumber\\
&=&\frac{1}{2}\int d\tau~d^{d}x~a^{d-1}(\tau)\left[(\partial_{\tau}\Phi(\tau,\vec{x}))^2-(\partial_{i}\Phi(\tau,\vec{x}))^2+m^2a^2(\tau)\Phi^2(\tau,\vec{x})\right].\eea
After varying this action with respect to the field the equation of motion, which is the {\it Klein Gordon equation} in $(d+1)$ dimensional de Sitter space can be written as:
\begin{equation}
\label{KevinGarnett}
\left(\nabla ^2_{\rm dS}-m^2\right) \Phi(\tau,\vec{x}) =0
\end{equation}
where, $\nabla ^2_{\rm dS}$ is the D'Alembertian operator of $(d+1)$ dimensional de Sitter space which is defined as:
\be \nabla ^2_{\rm dS}:=\frac{1}{\sqrt{-g_{(d+1)}}}\partial_{\mu}\left(\sqrt{-g_{(d+1)}}~g^{\mu\nu}\partial_{\nu}\right)=\frac{1}{l^2}\left[(d-1)\tau\partial_{\tau}-\tau^2\left(\partial^2_{\tau}-\partial^2_{i}\right)\right]\stackrel{\tau\rightarrow 0}{=}\frac{1}{l^2}\left[(d-1)\tau\partial_{\tau}-\tau^2\partial^2_{\tau}\right],\ee
where we are interested in the solution of the above equation at late time scales, i.e. $\tau \to 0$. Consequently, the asymptotic behaviour of the {\it Klein Gordon equation} can be represented by the following equation of motion:
\begin{equation}
\left\{\frac{1}{l^2}\left[(d-1)\tau\partial_{\tau}-\tau^2\partial^2_{\tau}\right]-m^2\right\} \Phi(\tau,\vec{x}) =0
\end{equation}
and the solution to this equation is given by:
\begin{equation}
    \Phi (\tau,\vec{x}) \sim \mathcal{O}_{\Delta_+} (\vec{x})~ \tau^{\Delta_+} +  \mathcal{O}_{\Delta_-} (\vec{x})~ \tau^{\Delta_-} 
\end{equation}
where, $\mathcal{O}_{\Delta_+}(\vec{x})$ and $\mathcal{O}_{\Delta_+}(\vec{x})$ represent the boundary operators which are characterized by the scaling dimensions.: 
 \be \Delta_\pm =\frac{d}{2}\pm i \nu~~~~{\rm where}~~~\nu=\sqrt{(ml)^2-\left(\frac{d}{2}\right)^2}~~~{\rm with}~~~(ml)^2=\Delta_+ \Delta_-\ee  
 Now we are interested in computing the two point and of course there are litany of these, for instance, retarded, advanced, Hadamard, Feynman and so on, but these are all encoded in the Wightman function which we write below:
\begin{equation}
\mathcal{G}(X_1, X_2) = \langle 0 | \Phi (X_1) \Phi(X_2) | 0 \rangle
\end{equation}
and it also obeys the homogeneous {\it Klein Gordon equation} as represented by Eq~(\ref{KevinGarnett}):
\begin{equation}
\label{KevinGarnett}
\left(\nabla ^2_{\rm dS}-m^2\right) \mathcal{G}(X_1, X_2)  =0~.
\end{equation}
Here the D'Alembertian operator in terms of $\sigma$ coordinate in $(d+1)$ dimensional de Sitter space can be expressed as:
\be \nabla ^2_{\rm dS}=\frac{1}{l^2}\left[\left(\frac{d+1}{2}\right)(1-2\sigma_{\rm dS})\partial_{\sigma_{\rm dS}}-\sigma(\sigma_{\rm dS}-1)\partial_{\sigma_{\rm dS}}\right].\ee
and further substituting this back in the {\it Klein Gordon equation} we get:
\be \left\{\frac{1}{l^2}\left[\left(\frac{d+1}{2}\right)(1-2\sigma_{\rm dS})\partial_{\sigma_{\rm dS}}-\sigma(\sigma_{\rm dS}-1)\partial_{\sigma_{\rm dS}}\right]-m^2\right\}\mathcal{G}(X_1, X_2) =0\ee
Let's look at the full solution to the above equation:
\begin{equation}
\mathcal{G} (\sigma) = A~ \, _2F_1\left(\frac{d}{2}+i \nu ,\frac{d}{2}-i \nu ;\frac{d+1}{2};\sigma_{\rm dS} \right) + B~ \, _2F_1\left(\frac{d}{2}+i \nu ,\frac{d}{2}-i \nu ;\frac{d+1}{2};\sigma_{\rm dS}-1 \right)
\end{equation}
and this solution is sometimes known as $\alpha$ vacua where the arbitrary constants $A$ and $B$ are parametrized by:
\be A=\cosh2\alpha~{\cal N}(d,\nu),~~~B=\sinh2\alpha~{\cal N}(d,\nu),~~{\rm with}~~|A|^2-|B|^2=|{\cal N}(d,\nu)|^2.\ee
Setting the parameter $\alpha=0$ gives:
\be A={\cal N}(d,\nu),~~~B=0,\ee
which corresponds to the Euclidean false vacuum state, which is commonly known as Bunch-Davies vacuum.  It is important to note in particular that the
Green functions which verify a condition (commonly known as the Hadamard condition) behave on
the light-cone as in flat space for Bunch Davies or the Euclidean false vacuum state. On the other
hand, the Bunch Davies or the Euclidean false vacuum can also be physically interpreted as being
generated by an infinite time tracing operation from the condition that the energy scale of the quantum
mechanical fluctuations is much smaller than the characteristic scale in cosmology, which is the Hubble
scale. This quantum vacuum state possesses actually no quanta at the limiting asymptotic past infinity.
However, in the framework of quantum field theory of curved space time, there exists a huge class
of quantum mechanical vacuum states in the background De Sitter space time which are invariant
under all the $SO(1, d+1)$ isometries and commonly known as the $\alpha$-vacua. Here $\alpha$ is a real parameter
which forms a real parameter family of continuous numbers to describe the issometric classes of
invariant quantum vacuum state in De Sitter space. In a more technical sense, sometimes the $\alpha$ vacua
is characterized as the squeezed quantum vacuum state in the context of quantum field theory of
curved space time. It is also important to note that in the original version something called, $\alpha,\beta$ vacua or
{\it Motta-Allen (MA) vacua} is appearing which is CPT violating and here an additional real parameter $\beta$
is appearing in the phases in the definition of the quantum mechanical vacuum state. This phase factor
is responsible for the CPT violation. Once we switch off this phase factor by fixing $\beta = 0$, the one
can get back the CPT symmetry preserving quantum vacuum state in the present context. The $\alpha$
vacua and the Bunch Davies or Euclidean false vacuum are connected to each other via Bogoliubov
transformation. Especially, the $\alpha=0$ case corresponds to the Bunch Davies or Euclidean vacuum state
in which the Hadamard condition in the Green’s functions is satisfied. 
Additionally, the point to be
noted here that the Bunch-Davies or the Euclidean quantum vacuum state is actually representing a
zero-particle quantum mechanical state which is observed by a geodesic observer, which implies that
an observer who is in free fall in the expanding state is characterized by this vacuum state. Because
of this reason to explain the origin of quantum mechanical fluctuations appearing in the context
of cosmological perturbation theory in the inflationary models or during the particle production
phenomena the concept of Euclidean false quantum vacuum state is commonly used in primordial
cosmology literature.  

It is worth noting that the solution is singular when $\sigma_{\rm dS} =1$. One can fix the overall coefficient ${\cal N}(d,\nu)$ by necessitating that the singularity is the same as the short distance singularity in flat space \footnote{At small distances field is not sensitive to the de Sitter space and hence the singularity is the same as the one that appears in the propagators of Minkowski space.}. For the Minkowski flat space considering the short distance singular behaviour one can write down the expression for the Wightman function as:
\be\mathcal{G}_{\rm flat} (X_1,X_2)\approx \frac{1}{{\cal D}^{d-1}_{\rm Geo}(X_1,X_2)} \frac{\Gamma\left(\frac{d+1}{2}\right)}{2(d-1)\pi^{\frac{d+1}{2}}},\ee
where ${\cal D}_{\rm Geo}(X_1,X_2)$ represents the geodesic distance between the two points represented by $X_1$ and $X_2$. The relationship between the $\sigma$ coordinate with the geodesic distance ${\cal D}_{\rm Geo}(X_1,X_2)$ can be represented by the following relation:
\be \sigma_{\rm dS}:=\sigma_{\rm dS}(X_1,X_2)=\frac{1}{2}\left[1+\cos\left(\frac{{\cal D}_{\rm Geo}(X_1,X_2)}{l}\right)\right].\ee
Further in terms of the $\sigma$ parametrization the flat space Wightman function can be further recast as:
\be\mathcal{G}_{\rm flat} (\sigma_{\rm dS})\approx \frac{1}{l^{d-1}\left[{\rm cos}^{-1}(2\sigma_{\rm dS}-1)\right]^{d-1}} \frac{\Gamma\left(\frac{d+1}{2}\right)}{2(d-1)\pi^{\frac{d+1}{2}}}.\ee
Now we use the following expansion of the Hypergeometric function around a point $z = 1$, which is given by:
\bea &&{}_{2}F_{1}(\alpha,\beta;\gamma;y)=\left[\frac{\Gamma(\gamma-\alpha-\beta)\Gamma(\gamma)}{\Gamma(\gamma-\alpha)\Gamma(\gamma-\beta)}+{\cal O}(y-1)\right]\nonumber\\
&& ~~~~~~~~~-(1-y)^{\gamma-\alpha-\beta}\exp\left(2i\pi(\gamma-\alpha-\beta)\left[\frac{{\rm arg}(y-1)}{2\pi}\right]\right)\left[\frac{\Gamma(\alpha+\beta-\gamma)\Gamma(\gamma)}{\Gamma(\alpha)\Gamma(\beta)}+{\cal O}(y-1)\right].~~~~~~~~~~\eea
Using this property we will now try to understand the simplified structures of the two Hypergeometric functions that are appearing in the expression for the full solution of the Wightman function in presence of $\alpha$ vacua:
\bea &&\, _2F_1\left(\frac{d}{2}+i \nu ,\frac{d}{2}-i \nu ;\frac{d+1}{2};\sigma_{\rm dS} \right)=\left[\frac{\Gamma\left(\frac{1-d}{2}\right)\Gamma\left(\frac{1+d}{2}\right)}{\Gamma\left(\frac{1}{2}-i\nu\right)\Gamma\left(\frac{1}{2}-i\nu\right)}+{\cal O}(\sigma_{\rm dS}-1)\right]\nonumber\\
&& -(1-\sigma_{\rm dS})^{\left(\frac{1-d}{2}\right)}\exp\left(i\pi(1-d)\left[\frac{{\rm arg}(\sigma_{\rm dS}-1)}{2\pi}\right]\right)\left[\frac{\Gamma\left(\frac{d-1}{2}\right)\Gamma\left(\frac{d+1}{2}\right)}{\Gamma\left(\frac{d}{2}+i \nu\right)\Gamma\left(\frac{d}{2}-i \nu\right)}+{\cal O}(\sigma_{\rm dS}-1)\right]\nonumber\\
&&~~~~~~~~~~~~~~~~~~~~~~~~~~~~~~~~~~~~~~~~~~~~\approx\frac{\Gamma\left(\frac{d-1}{2}\right)\Gamma\left(\frac{d+1}{2}\right)}{\Gamma\left(\frac{d}{2}+i \nu\right)\Gamma\left(\frac{d}{2}-i \nu\right)}\left(\frac{2l}{{\cal D}_{\rm Geo}(X_1,X_2)}\right)^{d-1},~~~~~~\eea\bea
 \, _2F_1\left(\frac{d}{2}+i \nu ,\frac{d}{2}-i \nu ;\frac{d+1}{2};\sigma_{\rm dS}-1 \right)&=&\left[\frac{\Gamma\left(\frac{1-d}{2}\right)\Gamma\left(\frac{1+d}{2}\right)}{\Gamma\left(\frac{1}{2}-i\nu\right)\Gamma\left(\frac{1}{2}-i\nu\right)}+{\cal O}(\sigma_{\rm dS}-2)\right]\nonumber\\
&& -\sigma^{\left(\frac{1-d}{2}\right)}_{\rm dS}\exp\left(i\pi(1-d)\left[\frac{{\rm arg}(\sigma_{\rm dS}-2)}{2\pi}\right]\right)\nonumber\\
&&~~~~~~\times\left[\frac{\Gamma\left(\frac{d-1}{2}\right)\Gamma\left(\frac{d+1}{2}\right)}{\Gamma\left(\frac{d}{2}+i \nu\right)\Gamma\left(\frac{d}{2}-i \nu\right)}+{\cal O}(\sigma_{\rm dS}-2)\right]\nonumber\\
&\approx &\frac{\Gamma\left(\frac{d-1}{2}\right)\Gamma\left(\frac{d+1}{2}\right)}{\Gamma\left(\frac{d}{2}+i \nu\right)\Gamma\left(\frac{d}{2}-i \nu\right)}\left(\frac{2l}{{\cal D}_{\rm Geo}(X_1,X_2)}\right)^{d-1}.
\eea
which will fix the overall normalization factor as given by the following expression:
\be {\cal N}(d,\nu):=\frac{ \Gamma \left(\frac{d}{2}-i \nu \right) \Gamma \left(\frac{d}{2}+i \nu \right)}{l^{d-1}(4 \pi )^{\frac{d+1}{2}}\Gamma \left(\frac{d+1}{2}\right)}.\ee
 Finally, the utilizing the short distance singular behaviour in the asymptotic flat space the full solution in presence of the $\alpha$ vacua can be written as:
\bea
\mathcal{G} (\sigma_{\rm dS}) &=&\left(\frac{\Gamma \left(\frac{d}{2}-i \nu \right) \Gamma \left(\frac{d}{2}+i \nu \right)}{l^{d-1}(4 \pi )^{\frac{d+1}{2}}\Gamma \left(\frac{d+1}{2}\right)} \right) ~\nonumber\\
&& 
 \left[\cosh2\alpha~\, _2F_1\left(\frac{d}{2}+i \nu ,\frac{d}{2}-i \nu ;\frac{d+1}{2};\sigma_{\rm dS} \right)+\sinh2\alpha~ \, _2F_1\left(\frac{d}{2}+i \nu ,\frac{d}{2}-i \nu ;\frac{d+1}{2};\sigma_{\rm dS}-1 \right)\right],\nonumber\\
 && \eea
 and as a special case for Bunch Davies vacua by setting $\alpha=0$ we get:
\bea  \mathcal{G} (\sigma_{\rm dS}) =\left(\frac{ \Gamma \left(\frac{d}{2}-i \nu \right) \Gamma \left(\frac{d}{2}+i \nu \right)}{l^{d-1}(4 \pi )^{\frac{d+1}{2}}\Gamma \left(\frac{d+1}{2}\right)} \right) ~\, _2F_1\left(\frac{d}{2}+i \nu ,\frac{d}{2}-i \nu ;\frac{d+1}{2};\sigma_{\rm dS} \right).\eea
    \begin{figure*}[htb]
    	\centering
    	{
    		\includegraphics[width=15.5cm,height=9.5cm] {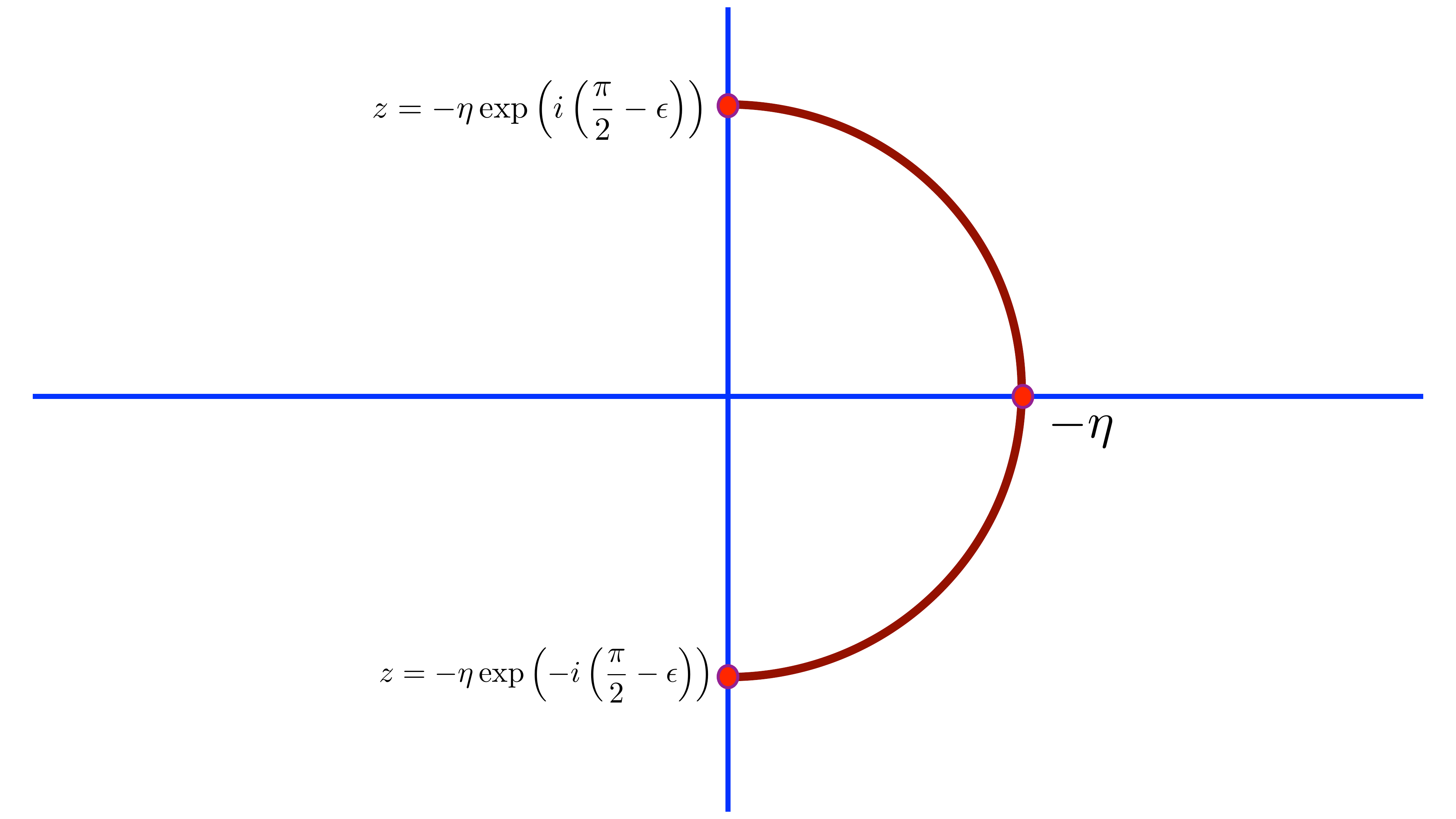}
    	}
    	\caption[Optional caption for list of figures]{Representative diagram showing analytic continuation from Anti de Sitter (AdS) to de Sitter (dS) space.} 
    	\label{Analytic}
    \end{figure*}
The hypergeometric function has a singularity at the short distance $\sigma_{\rm dS} =1$ and a branch cut for $1 < \sigma_{\rm dS} < \infty$. The singularity occurs when the two points become time like separated. We need an $i \epsilon$ prescription to go around the singularity in the complex plane for the flat de Sitter slicing. We have two options which can be written as:
\begin{equation}
\sigma^\pm_{\rm dS} =1+ \frac{\left(\tau _1- \tau _2\pm\frac{1}{2} i \epsilon \right){}^2-x_{12}^2}{4 \tau _1 \tau _2}
\end{equation}
where,
\begin{eqnarray}
  \mathcal{G}_{+-}(X_1, X_2) &=& \langle 0 | \hat{\Phi} (X_2) \hat{\Phi}(X_1) | 0 \rangle \nonumber\\
&=& \mathcal{G} (\sigma^+_{\rm dS})\nonumber\\
&=&\left(\frac{ \Gamma \left(\frac{d}{2}-i \nu \right) \Gamma \left(\frac{d}{2}+i \nu \right)}{l^{d-1}(4 \pi )^{\frac{d+1}{2}}\Gamma \left(\frac{d+1}{2}\right)} \right) ~\, _2F_1\left(\frac{d}{2}+i \nu ,\frac{d}{2}-i \nu ;\frac{d+1}{2};\sigma^+_{\rm dS} \right) , \\
\mathcal{G}_{-+}(X_1, X_2) &=& \langle 0 | \hat{\Phi}(X_1) \hat{\Phi}(X_2) | 0 \rangle\nonumber\\
&=& \mathcal{G} (\sigma^-_{\rm dS})\nonumber\\
&=&\left(\frac{ \Gamma \left(\frac{d}{2}-i \nu \right) \Gamma \left(\frac{d}{2}+i \nu \right)}{l^{d-1}(4 \pi )^{\frac{d+1}{2}}\Gamma \left(\frac{d+1}{2}\right)} \right) ~\, _2F_1\left(\frac{d}{2}+i \nu ,\frac{d}{2}-i \nu ;\frac{d+1}{2};\sigma^-_{\rm dS} \right)  
\end{eqnarray}

where the $-+$ and $+-$ corresponds to the analytic contributions appearing in $Y_1=\mp i X_1$ and $Y_2=\pm i X_2$ respectively. 

Using this information, one can further compute the expressions for the time-ordered and anti-time-ordered Wightman functions in the presence of $\alpha$ vacua in the present context, which are given by the following expressions:
\bea \langle 0 |{\cal T}\left( \hat{\Phi} (X_1) \hat{\Phi}(X_2)\right) | 0 \rangle&=&\theta(\tau_1-\tau_2)\mathcal{G}_{-+}(X_1, X_2)+\theta(\tau_2-\tau_1)\mathcal{G}_{+-}(X_1, X_2),\\
 \langle 0 |\bar{\cal T}\left( \hat{\Phi} (X_1) \hat{\Phi}(X_2)\right) | 0 \rangle&=&\theta(\tau_1-\tau_2)\mathcal{G}_{+-}(X_1, X_2)+\theta(\tau_2-\tau_1)\mathcal{G}_{-+}(X_1, X_2),\eea
  where the notation ${\cal T}$ and $\bar{\cal T}$ we represent the time-ordered and anti-time-ordered products. See refs. \cite{Bros:1995js,Spradlin:2001pw,Joung:2006gj,Baumann:2009ds,Anninos:2012qw,Akhmedov:2013vka,Allen:1985ux,Gibbons:1977mu,Bunch:1978yq,Burges:1984qm,Mottola:1984ar} for more details.
 
 In the time-dependent background geometry, it is further useful to use the {\it Schwinger-Keldysh formalism} or {\it in-in formalism} \cite{Chen:2017ryl} to compute the correlation functions and which are particularly very useful in the context of cosmology. In this formalism, to explicitly compute the fixed time expectation values in terms of correlation function one need to perform
a time-ordered integral which actually goes from the initial time to the time of interest at very late time scale, $\tau=\tau_0$~(one can choose $\tau_0=0$ for present day), and after that one need to further
perform an anti-time-ordered integral back to the initial time scale from which we have started doing the previous integral. This is in technical language is identified to be the 
{\it Schwinger-Keldysh contour} or {\it in-in contour}. The corresponding two point functions or the propagators with points that are considered along the different parts of the contour are given by the following expressions:
\bea \mathcal{G}_{++}(X_1, X_2)&=&\langle 0 |{\cal T}\left( \hat{\Phi} (X_1) \hat{\Phi}(X_2)\right) | 0 \rangle,\\
\mathcal{G}_{+-}(X_1, X_2)&=&\langle 0 |\hat{\Phi} (X_2) \hat{\Phi}(X_1) | 0 \rangle,\\
\mathcal{G}_{-+}(X_1, X_2)&=&\langle 0 |\hat{\Phi} (X_1) \hat{\Phi}(X_2) | 0 \rangle,\\
\mathcal{G}_{--}(X_1, X_2)&=&\langle 0 |\bar{\cal T}\left( \hat{\Phi} (X_1) \hat{\Phi}(X_2)\right) | 0 \rangle.\eea
One important thing we need to mention at the end is that all these results hold good for Bunch-Davies vacua which can be obtained by fixing $\alpha=0$, in that case only the definition of the quantum vacuum state will be changed and the important part is the $\alpha$ vacua and the Bunch Davies vacua are related via {\it Bogoliubov transformation}. In figure \ref{Analytic}, we have shown a representative diagram through which one can visualize the analytic continuation from AdS to dS space and this will be helpful to map the results obtained for AdS space to dS space.

\subsection{From the scalar field in AdS space}
\label{sec:2ads}

In the context of Anti De Sitter (AdS) space we have similar story just like de Sitter (dS) space. In this section we will establish the connection between these two space-times. To establish this analogy in AdS space, we consider a Harmonic function 
In AdS, the corresponding object is a Harmonic function $\mathcal{H}(Y_1, Y_2)$ which satisfy the following constraint condition:
\begin{equation}
\left( \nabla^2_{AdS} - m^2 \right) \mathcal{H}(Y_1, Y_2)=0\ .
\end{equation}
Here the D'Alembertian operator in terms of $\sigma$ coordinate in $(d+1)$ dimensional Anti de Sitter space can be expressed as:
\be \nabla ^2_{\rm AdS}=\frac{1}{l^2}\left[\left(\frac{d+1}{2}\right)(1-2\sigma_{\rm AdS})\partial_{\sigma_{\rm AdS}}-\sigma(\sigma_{\rm AdS}-1)\partial_{\sigma_{\rm AdS}}\right].\ee
and further substituting this back in the {\it Klein Gordon equation} we get:
\be \left\{\frac{1}{l^2}\left[\left(\frac{d+1}{2}\right)(1-2\sigma_{\rm AdS})\partial_{\sigma_{\rm AdS}}-\sigma(\sigma_{\rm AdS}-1)\partial_{\sigma_{\rm AdS}}\right]-m^2\right\}\mathcal{H}(\sigma_{\rm AdS}) =0\ee
And in the same vein as de Sitter, we can write down the full solution for the $SO(1,d+1)$ isommetric $\alpha$-vacua as given by:
\bea
\mathcal{H} (\sigma_{\rm AdS}) &=&\left(\frac{\Gamma \left(\frac{d}{2}-i \nu \right) \Gamma \left(\frac{d}{2}+i \nu \right)}{l^{d-1}(4 \pi )^{\frac{d+1}{2}}\Gamma \left(\frac{d+1}{2}\right)\Gamma(i\nu)\Gamma(-i\nu)} \right) ~\nonumber\\
&& 
~~~~~~~~~~~~ \left[\cosh2\alpha~\, _2F_1\left(\frac{d}{2}+i \nu ,\frac{d}{2}-i \nu ;\frac{d+1}{2};\sigma_{\rm AdS} \right)\right.\nonumber\\
&&\left.~~~~~~~~~~~~~~~~~~~~~~+\sinh2\alpha~ \, _2F_1\left(\frac{d}{2}+i \nu ,\frac{d}{2}-i \nu ;\frac{d+1}{2};\sigma_{\rm AdS}-1 \right)\right],~~~~~~
 \eea
 and as a special case for Bunch Davies vacua by setting $\alpha=0$ we get:
\bea
\mathcal{H} (\sigma_{\rm AdS}) =\left(\frac{\Gamma \left(\frac{d}{2}-i \nu \right) \Gamma \left(\frac{d}{2}+i \nu \right)}{l^{d-1}(4 \pi )^{\frac{d+1}{2}}\Gamma \left(\frac{d+1}{2}\right)\Gamma(i\nu)\Gamma(-i\nu)} \right) ~
\, _2F_1\left(\frac{d}{2}+i \nu ,\frac{d}{2}-i \nu ;\frac{d+1}{2};\sigma_{\rm AdS} \right).~~~~ \eea
While, there are at the face of it, it looks similar to the $G(\sigma\sigma_{\rm dS})$, besides the overall difference in the factor of $\Gamma(i \nu) \Gamma(-i \nu)$ appearing in the denominator, the short distance limit occurs when $\sigma_{{AdS}} \rightarrow 0$. Curiously compared to the de Sitter case, this is not singular. However, we can map the Harmonic function in AdS to the Wightman function  through the following sets of analytic continuation. We can write,
\begin{align}
\mathcal{G}_{+-}(X_1, X_2) &= \mathcal{H} (-i X_1, i X_2) \times \Gamma( i \nu) \Gamma(- i \nu)\\
\mathcal{G}_{-+}(X_1, X_2) &= \mathcal{H} (i X_1, -i X_2) \times \Gamma(i \nu) \Gamma(-i \nu)
\end{align}
Further this Harmonic function in AdS which represents the teh bulk to bulk propagator can be written in terms of \emph{split representation} by the following integral transformation:
\begin{equation}
  \mathcal{H} (Y_1,Y_2):= \frac{\nu ^2} {\pi} \int dP~K_{\Delta _+} \left(Y_1,P\right) K_{\Delta _-}\left(Y_2,P\right) \, 
\end{equation}
where, two bulk to boundary propagators $K_{\Delta _+} \left(Y_1,P\right)$ and $ K_{\Delta _-}\left(Y_2,P\right)$ are integrated over the boundary point having the coordinate $P$. The bulk to boundary propagator, the basic ingredient required to compute AdS correlators is given by:
\bea
K_{\Delta_{+}}(Y_1,P)&=&\frac{\mathcal{C}_{\Delta_{+}} }{(-2 P \cdot Y_1)^{\Delta_{+}}},\\
K_{\Delta_{-}}(Y_2,P)&=&\frac{\mathcal{C}_{\Delta_{-}} }{(-2 P \cdot Y_2)^{\Delta_{-}}}\eea
where we define $\mathcal{C}_\Delta$ by the following expression:
\bea
\mathcal{C}_{\Delta_{+}} &=& \frac{ \Gamma({\Delta_{+}})}{2 \pi^h \Gamma \left({\Delta_{+}}-h+1 \right)}\, \\
\mathcal{C}_{\Delta_{-}} &=& \frac{ \Gamma({\Delta_{-}})}{2 \pi^h \Gamma \left({\Delta_{-}}-h+1 \right)}\, \ \ \ \ \ \ \ \text{where}\,  \ \ \ \ \ \ \ h:= \frac{d}{2}
\eea
Armed with all these ingredients, after analytically continuing in the dS space we get the following expression for the bulk to bulk propagator for the dS space:

\begin{equation}
  \mathcal{G}(X_1,X_2):=  \int dP~\mathcal{K}_{\Delta _+} \left(\mp i X_1,P\right) \mathcal{K}_{\Delta _-} \left(\pm i X_2,P\right) \, ~~~{\rm where}~~\Delta_{\pm}:=\frac{d}{2}\pm i\nu,
\end{equation}
where, two bulk to boundary propagators after analytical continuation from $Y_1\rightarrow \mp i X_1$ and $Y_2\rightarrow \pm i X_2$, are represented by, $\mathcal{K}_{\Delta _+} \left(\mp i X_1,P\right) $ and $\mathcal{K}_{\Delta _-} \left(\pm i X_2,P\right)$:
\bea {\cal K}_{\Delta _{+}}\left(\mp i X_1,P\right) :&=&\frac{\Gamma\left(\Delta _{+}-h+1\right)}{\sqrt{\pi}}~K_{\Delta _{+}}\left(\mp i X_1,P\right),\\
{\cal K}_{\Delta _{-}}\left(\pm i X_2,P\right) :&=&\frac{\Gamma\left(\Delta _{-}-h+1\right)}{\sqrt{\pi}}~K_{\Delta _{-}}\left(\pm i X_2,P\right).\eea
where both of them are integrated over the boundary coordinate $P$. The corresponding bulk to boundary propagator, the basic ingredient required to compute AdS correlators is given by:
\bea
&& {\cal K}_{\Delta_{+}} (Y_1,P)=\frac{\mathcal{C}_\Delta }{(-2 P \cdot Y_1)^{\Delta}}\longrightarrow {\cal K}_{\Delta_{+}} (\mp i X_1,P)=\frac{\mathcal{C}_{\Delta_{+}} }{(-2 P \cdot (\mp i X_1))^{\Delta_{+}}},\\
&& {\cal K}_{\Delta_{-}} (Y_2,P)=\frac{\mathcal{C}_{\Delta_{-}}  }{(-2 P \cdot Y_1)^{{\Delta_{-}} }}\longrightarrow {\cal K}_{\Delta_{-}}  (\pm i X_2,P)=\frac{\mathcal{C}_{\Delta_{-}}  }{(-2 P \cdot (\pm i X_2))^{{\Delta_{-}} }}.
\eea
Here it is important to note that the structure of $\mathcal{C}_{\Delta_{+}} $ and $\mathcal{C}_{\Delta_{-}} $, for both AdS and dS become exactly same because this factor is only dependent on conformal dimensions of the operators as well as the spatial dimension $d$, which are used to compute the two point functions in the both AdS and dS space. But this quantity is not at all dependent on the coordinate before (for AdS) and after (for dS) analytical continuation in the coordinate from the AdS to dS space. In the present discussion, the {\it split representation} have been used as an instrumental technique, particularly in the evaluation of bulk
Witten diagrams in EAdS and is suitable to obtain the {\it Conformal
Partial Wave decomposition} of tree-level exchange in the bulk Witten diagrams \cite{Costa:2014kfa,Hartman:2006dy,Giombi:2011ya,Bekaert:2014cea,Chen:2017yia,Tamaoka:2017jce,Giombi:2017hpr,Sleight:2017cax,Nishida:2018opl,Costa:2018mcg,Carmi:2018qzm,Zhou:2018sfz,Jepsen:2019svc,Leonhardt:2003qu,Costa:2011dw,Joung:2011ww,Joung:2012fv}. This further helps further to factorise
the Harmonic functions in EAdS into an integrated product of three-point Witten diagrams. In this
paper our prime objective is to explicitly show that the {\it split representation} is also very useful mathematical trick in the context of de Sitter space, where at the late-time scale
the tree-level exchange diagrams in the context of $dS_{d+1 }$can be obtained from existing results for $EAdS_{d+1}$ particularly for
three-point bulk Witten diagrams just making use of the analytic continuation.

\section{Three point function }
\subsection{From the scalar field in AdS space}
\label{sec:3ads}

In this section our prime objective is to look back to the computation of the three point function for scalar fields in the background of Anti-de Sitter space time. For this purpose we will start with the $O(3)$ theory of a scalar fields $\Phi_i$ of mass $m_i$ having identical cubic interaction strength $g$ is represented by the following action:
\bea S&=&\int d^{d+1}x~\sqrt{-g_{(d+1)}}~\sum^{3}_{i=1}\left[-\frac{1}{2}(\partial\Phi_i)^2+\frac{m^2_i}{2}\Phi^2_i+\frac{g}{3!}\Phi^3_i\right].\eea
In this context, the conformal dimension of the operators dual to $\Phi_i$ is given by:
\bea \Delta_i:=h\pm \sqrt{h^2+m^2_i}=\frac{d}{2}\left(1\pm \sqrt{1+\left(\frac{2m_i}{d}\right)^2}\right)~~~~~{\rm where} ~~h=\frac{d}{2}.\eea

Then the corresponding  three point function for this $O(3)$ scalar field theory is composed of three bulk to boundary propagator, and can be represented by the following equation in AdS space:
\be
\langle \mathcal{O}_1(P_1)\mathcal{O}_2(P_2) \mathcal{O}_3(P_3) \rangle = g \int_0^\infty dX~~K_{\Delta_1} (X, P_1) K_{\Delta_2} (X, P_2) K_{\Delta_3} (X, P_3)
\ee	
Further to simplify the right hand side of the above mentioned bulk boundary integrals we use {\it Schwinger parametrization}, for which we get the following result:
\begin{equation}
\langle \mathcal{O}_1(P_1)\mathcal{O}_2(P_2) \mathcal{O}_3(P_3) \rangle =g ~\mathcal{E}_3 \int_0^ \infty \prod_{i=1}^3 \frac{dt_i}{t_i} t^{\Delta_i} \int_{\text{AdS}} dX~ e^{2 (t_1 P_1 + t_2 P_2 + t_3 P_3) \cdot X} 
\end{equation} where, the factor $\mathcal{E}_3$ is given by the following expression:
\be 
\mathcal{E}_3= \prod_{j=1}^3 \frac{C_j} {\Gamma{(\Delta_j)}}=\prod_{j=1}^3 \frac{1} {2\pi^{h}\Gamma{(\Delta_j-h+1)}}.\ee
Here the bulk integral can be expressed as, which is our further aim to evaluate in the present context:
\begin{equation}
 \mathcal{I}_{\bf Bulk}\sim \int_{\text{AdS}} dX  (-2P_i\cdot X)^{-\Delta_i} \sim \int_0^\infty \prod_i \frac{dt_i}{t_i} t_i^{\Delta_i} \int_{\text{AdS}} dX~ e^{2T\cdot X} 
 \end{equation}
 where we have introduced a new short-hand factor $T$, which is defined as:
 \be T:=\sum^{3}_{i=1}t_iP_i.\ee
Here before performing the above mentioned integral it is important to note that the bulk point was parametrized in the following way:
\begin{equation}
X= \frac{1}{x_0} \left( \frac{x_0^2 + x^2+1}{2},  \frac{x_0^2 + x^2-1}{2}, x^\mu \right)
\end{equation}
Hence, it is very straightforward to show from the above mentioned expression for the bulk integral that one can easily obtain the following simplified result:
\begin{equation}
 \int_{\text{AdS}} dX~ e^{2T\cdot X} = \pi^{d/2} \int_0^\infty \frac{dx_0}{x_0}~ x_0^{-d/2}~ e^{-x_0+T^2/x_0} 
 \end{equation}
Further rescaling $t_i \to t_i\sqrt{ x_0}$, one can integrate over $x_0$ and obtain the following result for the bulk integral:
\begin{equation}
\mathcal{I}_{\bf Bulk} = \pi^{d/2}~ \Gamma \left( \frac{\displaystyle \sum^{3}_{i=1}\Delta_i -d}{2} \right) \int_0^\infty \prod^{3}_{i=1} \frac{dt_i}{t_i}~ t_i^{\Delta_i}~ e^{T^2} 
 \end{equation}
After using the above mentioned formalism the three point function can be further simplified as:
	\begin{eqnarray}
	\langle \mathcal{O}_1(P_1) \mathcal{O}_2(P_2) \mathcal{O}_3(P_3) \rangle & = & \frac{g}{2} \pi^h\mathcal{E}_3 \Gamma\left( \frac{\displaystyle \sum_{i=1}^3 \Delta_i- 2h }{2}\right) \int_0^ \infty \prod_{i=1}^3 \frac{dt_i}{t_i} t^{\Delta_i} e^{(t_1 P_1 + t_2 P_2 + t_3 P_3)^2} \nonumber \\
	&=&\frac{g}{2} \pi^h\mathcal{E}_3 \Gamma\left( \frac{\displaystyle \sum_{i=1}^3 \Delta_i- 2h }{2}\right)\int_0^ \infty \prod_{i=1}^3 \frac{dt_i}{t_i} t^{\Delta_i} e^{ - (t_1 t_2 P_{12} + t_2 t_3 P_{23} + t_1t_3 P_{13})}.\nonumber\\
	&&
	\end{eqnarray}
	where we define, $h=d/2$ in the present context. For the further computation, it is important to note that, we have earlier defined, $ P_{ij} = - 2 P_i \cdot P_j$. Also, we want to change the variables to do the computations in a simplest fashion, i.e, 
\begin{align}
t_1 = \sqrt{\frac{m_3 m_2} {m_1}}, \qquad  t_2 = \sqrt{\frac{m_1 m_3} {m_2}}, \qquad t_3 = \sqrt{\frac{m_1 m_2} {m_3}}~.
\end{align}
Further implementing these change of variables the three point function can be simplified as:
\begin{equation}
\langle \mathcal{O}_1(P_1) \mathcal{O}_2(P_2) \mathcal{O}_3(P_3) \rangle =\frac{g}{2} \pi^h\mathcal{E}_3 \Gamma\left( \frac{\displaystyle\sum_{i=1}^3 \Delta_i- 2h }{2}\right) \int_0^ \infty \prod_{i=1}^3 \frac{dm_i}{m_i} m_i^{\delta_{jk}} e^{-m_i P_{jk}}
\end{equation}
where, it is important to note that, $i=1$ and $j,k= 2, 3$ and so on. We also define the following quantities, which will be extremely useful for the further simplification, and as  given by:
\begin{equation}
\delta_{12} = \frac{\Delta_1+ \Delta_2- \Delta_3}{2}, \qquad \delta_{23} = \frac{\Delta_2+ \Delta_3- \Delta_1}{2},  \qquad \delta_{13} = \frac{\Delta_1+ \Delta_3- \Delta_2}{2}.
\end{equation}
This change in variables makes the integration crisp and clear, which can be recast in the following form:
\begin{equation}
\langle \mathcal{O}_1(P_1) \mathcal{O}_2(P_2) \mathcal{O}_3(P_3) \rangle = \frac{g}{2} \pi^h\mathcal{E}_3 \Gamma\left( \frac{\displaystyle \sum_{i=1}^3 \Delta_i- 2h }{2}\right) \prod_{i <j}^3 \Gamma({\delta_{ij}}) P_{ij}^{-\delta_{ij}}
\end{equation}
In this specific example of AdS space time, the three point {\it Mellin Barnes amplitude} is given by the following expression:
\bea {\cal M}_3:&=&g~\Gamma\left( \frac{\displaystyle\sum_{i=1}^3 \Delta_i- 2h }{2}\right)\nonumber\\
&=&g~\Gamma\left(\frac{d}{4} \left[\displaystyle\sum_{i=1}^3 \left(1\pm \sqrt{1+\left(\frac{2m_i}{d}\right)^2}\right)- 2\right]\right).\eea
One can further consider few special cases further, to study the various outcomes of the three point function and the related {\it Mellin Barnes amplitude} derived in this section for the AdS space time. First of all one can consider computing the three point function and the related {\it Mellin Barnes amplitude} from three identical scalar fields having the same mass parameter, $m$. In that case the expression the conformal dimension of each individual operators participating in the three point function computation will be further simplified by replacing $m_i$ with $m$. This further implies that the conformal dimension of the there operators are identical in this particular case and given by:
\bea \Delta:=\Delta_1=\Delta_2=\Delta_3=h\pm \sqrt{h^2+m^2}.\eea
In this particular case, we further have the following simplifications:
\begin{equation}
\delta:=\delta_{12} = \delta_{23} =\delta_{13} = \frac{\Delta}{2}.
\end{equation}
In this case, the three point function can be further simplified as:
\begin{equation}
\langle \mathcal{O}_1(P_1) \mathcal{O}_2(P_2) \mathcal{O}_3(P_3) \rangle = \frac{g}{2} \pi^h\mathcal{E}_3 \Gamma\left( \frac{\displaystyle 3\Delta- 2h }{2}\right) \Gamma\left(\frac{\Delta}{2}\right)~\prod_{i <j}^3  P_{ij}^{-\frac{\Delta}{2}},
\end{equation}
and the corresponding {\it Mellin Barnes amplitude} is given by the following expression:
\bea {\cal M}_3:&=&g~\Gamma\left( \frac{\displaystyle 3\Delta- 2h }{2}\right)\nonumber\\
&=&g~\Gamma\left(\frac{d}{4} \left[\displaystyle 3\left(1\pm \sqrt{1+\left(\frac{2m}{d}\right)^2}\right)- 2\right]\right).\eea
Particularly in $d=3$ the conformal dimension of the identical operators can be written as:
\bea \Delta=\frac{3}{2}\left(1\pm \sqrt{1+\left(\frac{2m}{3}\right)^2}\right)~~~~{\rm with}~~~~h=\frac{3}{2},\eea
which will be, $\Delta=3,0$ for the massless scalars. Here one can further check that for $d=3$ we get:
\begin{eqnarray}
&&\Delta=0~~~\longrightarrow~~~{\cal M}_3=g~\Gamma\left(-\frac{3}{2}\right)=\frac{4g\sqrt{\pi}}{3},~~~~\langle \mathcal{O}_1(P_1) \mathcal{O}_2(P_2) \mathcal{O}_3(P_3) \rangle \rightarrow \infty, \\
&&\Delta=3~~~\longrightarrow~~~{\cal M}_3=g~\Gamma\left(3\right)=2~g,~~~~~~~~~~~~~\langle \mathcal{O}_1(P_1) \mathcal{O}_2(P_2) \mathcal{O}_3(P_3) \rangle = \frac{g}{2}\pi^{2}\mathcal{E}_3~\prod_{i <j}^3  P_{ij}^{-\frac{3}{2}},~~~~~~~~~~
\end{eqnarray}
 where, the factor $\mathcal{E}_3$ is given by for the massless scalar field by the following expression:
\be 
\mathcal{E}_3= \prod_{n=1}^3 \frac{ 1}{2 \pi^{\frac{3}{2}} \Gamma \left(\frac{5}{2} \right)}= \frac{ 8}{27\pi^6}.\ee
Here one additional remark we want to make before going to the next section is that when we consider massless scalar theories we always get one conformal dimension, $\Delta=0$ which give rise to divergent three point function. So the above mentioned pathology is not the outcome of only $d=3$, but will valid for any arbitrary spatial dimension with massless scalar fields. This can be demonstrated for any arbitrary $d$ with massless scalar fields as following:
\bea \Delta=h\pm h=2h,0=d,0~~~{\rm with~~} h=\frac{d}{2}.\eea
Consequently, we get the following expressions for the {\it Mellin Barnes amplitude} and the three point function:
\begin{eqnarray}
&&\Delta=0~~~\longrightarrow~~~{\cal M}_3=g~\Gamma\left(-\frac{d}{2}\right),~~~~\langle \mathcal{O}_1(P_1) \mathcal{O}_2(P_2) \mathcal{O}_3(P_3) \rangle \rightarrow \infty, \\
&&\Delta=d~~~\longrightarrow~~~{\cal M}_3=g~\Gamma\left(d\right),~~\langle \mathcal{O}_1(P_1) \mathcal{O}_2(P_2) \mathcal{O}_3(P_3) \rangle = \frac{g}{2}\pi^{d/2}\Gamma(d)\Gamma\left(\frac{d}{2}\right)\mathcal{E}_3~\prod_{i <j}^3  P_{ij}^{-\frac{d}{2}},~~~~~~~~~~
\end{eqnarray}
 where, the factor $\mathcal{E}_3$ is given by for the massless scalar field by the following expression:
\be 
\mathcal{E}_3= \prod_{n=1}^3 \frac{ 1}{2 \pi^{\frac{d}{2}} \Gamma \left(1+\frac{d}{2} \right)}= \frac{ 1}{8 \pi^{\frac{3d}{2}}  \Gamma^3 \left(1+\frac{d}{2} \right)}.\ee

\subsection{From the scalar field in dS space}
\label{sec:3ds}

In this section our prime objective is to compute of the three point function for scalar fields in the background of de Sitter space time. For this purpose we will start with the $O(3)$ theory of a scalar fields $\Phi_q$ of mass $m_q$ having identical cubic interaction strength $g$ is represented by the following action:
\bea S&=&\int d^{d+1}x~\sqrt{-g_{(d+1)}}~\sum^{3}_{q=1}\left[-\frac{1}{2}(\partial\Phi_q)^2+\frac{m^2_q}{2}\Phi^2_q+\frac{g}{3!}\Phi^3_q\right]\nonumber\\
&=&\frac{1}{2}\int d\tau~d^{d}x~a^{d-1}(\tau)\sum^{3}_{q=1}\left[(\partial_{\tau}\Phi_q(\tau,\vec{x}))^2-(\partial_{i}\Phi_q(\tau,\vec{x}))^2+m^2_q a^2(\tau)\Phi^2_q(\tau,\vec{x})\right.\nonumber\\
&&\left.~~~~~~~~~~~~~~~~~~~~~~~~~~~~~~~~~~~~~~~~~~~~~~~~~~~~~~~~~~~~~~~~~~~~~~~~~~~~~~~~+\frac{g}{3}a^2(\tau)\Phi^3_q(\tau,\vec{x})\right].~~~~~~~~~~~~\eea

Then the corresponding  three point function for this $O(3)$ scalar field theory is composed of three bulk to boundary propagator, and can be represented by the following equation in dS space, after analytically continuing from AdS space as:
\be
\langle \mathcal{O}_1(P_1)\mathcal{O}_2(P_2) \mathcal{O}_3(P_3) \rangle = \langle\mathcal{O}_1(P_1)\mathcal{O}_2(P_2) \mathcal{O}_3(P_3) \rangle_{+}+\langle\mathcal{O}_1(P_1)\mathcal{O}_2(P_2) \mathcal{O}_3(P_3) \rangle_{-},
\ee
where each of the individual contributions appearing in the above mentioned expression are appended below:	
\be
\langle \mathcal{O}_1(P_1)\mathcal{O}_2(P_2) \mathcal{O}_3(P_3) \rangle_{+} = g \int_0^\infty dX~~{\cal K}_{\Delta^{+}_{1}} (\mp iX, P_1) ~{\cal K}_{\Delta^{+}_{2}} (\mp iX, P_2)~ {\cal K}_{\Delta^{+}_{3}} (\mp iX, P_3)
\ee
\be
\langle \mathcal{O}_1(P_1)\mathcal{O}_2(P_2) \mathcal{O}_3(P_3) \rangle_{-} = g \int_0^\infty dX~~{\cal K}_{\Delta^{-}_{1}} (\pm iX, P_1) ~{\cal K}_{\Delta^{-}_{2}} (\pm iX, P_2)~ {\cal K}_{\Delta^{-}_{3}} (\pm iX, P_3)
\ee		
In this context, the conformal dimension of the operators dual to $\Phi_n$ is given by:
\bea \Delta^{\pm}_n:=\frac{d}{2}\pm i\nu_n~~~{\rm where}~~\nu_n=\sqrt{(m_nl)^2-\left(\frac{d}{2}\right)^2}~~{\rm with}~~(m_nl)^2=\Delta^{+}_{n}\Delta^{-}_{n},~~\forall~n=1,2,3.~~~~\eea
In this context, bulk to boundary propagators in dS space after analytically continuing from AdS space we get:
\bea {\cal K}_{\Delta^{+}_n}\left(\mp i X_1,P\right) :&=&\frac{\Gamma\left(\Delta^{+}_n-\frac{d}{2}+1\right)}{\sqrt{\pi}}~K_{\Delta^{+}_n}\left(\mp i X_1,P\right)=\frac{\mathcal{C}_{\Delta^{+}_n} }{(-2 P \cdot (\mp i X_1))^{\Delta^{+}_n}},~~\forall~n=1,2,3,\\
{\cal K}_{\Delta^{-}_n}\left(\pm i X_2,P\right) :&=&\frac{\Gamma\left(\Delta^{-}_n-\frac{d}{2}+1\right)}{\sqrt{\pi}}~K_{\Delta^{-}_n}\left(\pm i X_2,P\right)=\frac{\mathcal{C}_{\Delta^{-}_n} }{(-2 P \cdot (\mp i X_1))^{\Delta^{-}_n}},~~\forall~n=1,2,3,~~~~~~~~~\eea
where we define $C_{\Delta^{\pm}_n}~~\forall~~n=1,2,3$ by the following expression:
\bea
\mathcal{C}_{\Delta^{\pm}_n} &=& \frac{ \Gamma({\Delta^{\pm}_n})}{2 \pi^{\frac{d}{2}} \Gamma \left({\Delta^{\pm}_n}-\frac{d}{2}+1 \right)}\,.
\eea
Further to simplify the right hand side of the above mentioned bulk boundary integrals we use {\it Schwinger parametrization}, for which we get the following result of the two individual contribution of the scalar three point function in dS space:
\begin{equation}
\langle \mathcal{O}_1(P_1)\mathcal{O}_2(P_2) \mathcal{O}_3(P_3) \rangle_{+} =g ~\mathcal{E}^{+}_3 \int_0^ \infty \prod_{l=1}^3 \frac{dt_l}{t_l} t^{\Delta^{+}_l}_l \int_{\text{dS}} dX~ e^{\mp 2i (t_1 P_1 + t_2 P_2 + t_3 P_3) \cdot X} 
\end{equation} 
\begin{equation}
\langle \mathcal{O}_1(P_1)\mathcal{O}_2(P_2) \mathcal{O}_3(P_3) \rangle_{-} =g ~\mathcal{E}^{-}_3 \int_0^ \infty \prod_{j=1}^3 \frac{ds_j}{s_j} s^{\Delta^{-}_j}_j \int_{\text{dS}} dX~ e^{\pm 2i (s_1 P_1 + s_2 P_2 + s_3 P_3) \cdot X} 
\end{equation} 
where, the factors $\mathcal{E}^{+}_3$ and $\mathcal{E}^{-}_3$ are given by the following expressions:
\be 
\mathcal{E}^{\pm}_3= \prod_{n=1}^3 \frac{C_{\Delta^{\pm}_n}} {\Gamma{(\Delta^{\pm}_n)}}= \prod_{n=1}^3 \frac{C_{\Delta^{\pm}_n}} {\Gamma{\left(\frac{d}{2}\pm i\nu_n\right)}}.\ee

Here the bulk integral can be expressed as, which is our further aim to evaluate in the present context:
\begin{equation}
 \mathcal{I}^{+}_{\bf Bulk}\sim\int_{\text{dS}} dX (-2P_n\cdot (\mp iX))^{-\Delta^{+}_n}\sim \int_0^\infty \prod^{3}_{n=1} \frac{dt_n}{t_n} t_n^{\Delta^{+}_n} \int_{\text{dS}} dX~ e^{\mp2iT\cdot X} 
 \end{equation}
 \begin{equation}
 \mathcal{I}^{-}_{\bf Bulk}\sim \int_{\text{dS}} dX (-2P_n\cdot (\pm iX))^{-\Delta^{-}_n}\sim\int_0^\infty \prod^{3}_{n=1} \frac{ds_n}{s_n} s_n^{\Delta^{-}_n} \int_{\text{dS}} dX~ e^{\pm2iS\cdot X}  
 \end{equation}
 where we have introduced a new short-hand factor $T$ and $S$, which are defined as:
 \be T:=\sum^{3}_{i=1}t_iP_i,~~~~~S:=\sum^{3}_{j=1}s_jP_j.\ee
Here before performing the above mentioned integral it is important to note that the bulk point was parametrized in the following way:
\begin{equation}
X= \frac{1}{x_0} \left( \frac{x_0^2 + x^2+1}{2},  \frac{x_0^2 + x^2-1}{2}, x^\mu \right)
\end{equation}
 Hence, it is very straightforward to show from the above mentioned expression for the bulk integral in the dS space that one can easily obtain the following simplified result:
\begin{equation}
 \int_{\text{dS}} dX~ e^{\mp2iT\cdot X} = \pi^{d/2} \int_0^\infty \frac{dx_0}{x_0}~ x_0^{-d/2}~ e^{-x_0-T^2/x_0} 
 \end{equation}
 \begin{equation}
 \int_{\text{dS}} dX~ e^{\pm2iS\cdot X} = \pi^{d/2} \int_0^\infty \frac{dx_0}{x_0}~ x_0^{-d/2}~ e^{-x_0-S^2/x_0} 
 \end{equation}
Further rescaling $t_i \to t_i\sqrt{ x_0}$ and $s_j\to s_j\sqrt{x_0}$, one can integrate over $x_0$ and obtain the following results for the bulk  integrals in the dS space:
\bea
\mathcal{I}^{+}_{\bf Bulk} = \pi^{d/2}~ \Gamma \left( \frac{\displaystyle \sum^{3}_{n=1}\Delta^{+}_n -d}{2} \right) \int_0^\infty \prod^{3}_{i=1} \frac{dt_i}{t_i}~ t_i^{\Delta^{+}_i}~ e^{-T^2} 
 \\
\mathcal{I}^{-}_{\bf Bulk} = \pi^{d/2}~ \Gamma \left( \frac{\displaystyle \sum^{3}_{n=1}\Delta^{-}_n -d}{2} \right) \int_0^\infty \prod^{3}_{j=1} \frac{ds_j}{s_j}~ s_i^{\Delta^{-}_j}~ e^{-S^2} 
 \eea
After using the above mentioned formalism the three point function can be further simplified as:
	\begin{eqnarray}
	\langle \mathcal{O}_1(P_1) \mathcal{O}_2(P_2) \mathcal{O}_3(P_3) \rangle_{+} & = & \frac{g}{2} \pi^{\frac{d}{2}}\mathcal{E}^{+}_3 \Gamma\left( \frac{\displaystyle \sum_{n=1}^3 \Delta^{+}_n- d }{2}\right) \int_0^ \infty \prod_{i=1}^3 \frac{dt_i}{t_i} t^{\Delta^{+}_i} e^{-(t_1 P_1 + t_2 P_2 + t_3 P_3)^2}\nonumber \\
	&=&\frac{g}{2} \pi^{\frac{d}{2}}\mathcal{E}^{+}_3 \Gamma\left( \frac{\displaystyle \sum_{n=1}^3 \Delta^{+}_n- d }{2}\right)\int_0^ \infty \prod_{i=1}^3 \frac{dt_i}{t_i} t^{\Delta^{+}_i} e^{  (t_1 t_2 P_{12} + t_2 t_3 P_{23} + t_1t_3 P_{13})}.~~~~~
	\end{eqnarray} 
	and 
	\begin{eqnarray}
	\langle \mathcal{O}_1(P_1) \mathcal{O}_2(P_2) \mathcal{O}_3(P_3) \rangle_{-} & = & \frac{g}{2} \pi^{\frac{d}{2}}\mathcal{E}^{-}_3 \Gamma\left( \frac{\displaystyle \sum_{n=1}^3 \Delta^{-}_n- d }{2}\right) \int_0^ \infty \prod_{i=1}^3 \frac{ds_i}{s_i} s^{\Delta^{-}_i} e^{-(s_1 P_1 + s_2 P_2 + s_3 P_3)^2}\nonumber \\
	&=&\frac{g}{2} \pi^{\frac{d}{2}}\mathcal{E}^{-}_3 \Gamma\left( \frac{\displaystyle \sum_{n=1}^3 \Delta^{-}_n- d }{2}\right)\int_0^ \infty \prod_{i=1}^3 \frac{ds_i}{s_i} t^{\Delta^{-}_i} e^{  (s_1 s_2 P_{12} + s_2 s_3 P_{23} + s_1s_3 P_{13})}.\nonumber\\ 
	&&
	\end{eqnarray} 
	For the further computation, it is important to note that, we have earlier defined, $ P_{ij} = - 2 P_i \cdot P_j$. Also, we want to change the variables to do the computations in a simplest fashion, i.e, 
\begin{align}
t_1 = \sqrt{\frac{m_3 m_2} {m_1}}, \qquad  t_2 = \sqrt{\frac{m_1 m_3} {m_2}}, \qquad t_3 = \sqrt{\frac{m_1 m_2} {m_3}}~.
\end{align}
Further implementing these change of variables individual contributions of the scalar three point function can be simplified as:
\begin{equation}
\langle \mathcal{O}_1(P_1) \mathcal{O}_2(P_2) \mathcal{O}_3(P_3) \rangle_{+} =\frac{g}{2} \pi^{\frac{d}{2}}\mathcal{E}^{+}_3 \Gamma\left( \frac{\displaystyle\sum_{n=1}^3 \Delta^{+}_n- d }{2}\right) \int_0^ \infty \prod_{i=1}^3 \frac{dm_i}{m_i} m_i^{\delta^{+}_{jk}} e^{m_i P_{jk}}
\end{equation}
\begin{equation}
\langle \mathcal{O}_1(P_1) \mathcal{O}_2(P_2) \mathcal{O}_3(P_3) \rangle_{-} =\frac{g}{2} \pi^{\frac{d}{2}}\mathcal{E}^{-}_3 \Gamma\left( \frac{\displaystyle\sum_{n=1}^3 \Delta^{-}_n- d }{2}\right) \int_0^ \infty \prod_{i=1}^3 \frac{dm_i}{m_i} m_i^{\delta^{-}_{jk}} e^{m_i P_{jk}}
\end{equation} 
where, it is important to note that, $i=1$ and $j,k= 2, 3$ and so on. We also define the following quantities, which will be extremely useful for the further simplification, and as  given by:
\begin{equation}
\delta^{\pm}_{12} = \frac{\Delta^{\pm}_1+ \Delta^{\pm}_2- \Delta^{\pm}_3}{2}, \qquad \delta_{23} = \frac{\Delta^{\pm}_2+ \Delta^{\pm}_3- \Delta^{\pm}_1}{2},  \qquad \delta^{\pm}_{13} = \frac{\Delta^{\pm}_1+ \Delta^{\pm}_3- \Delta^{\pm}_2}{2}.
\end{equation}
This change in variables makes the integration crisp and clear, which can be recast in the following form:
\begin{equation}
\langle \mathcal{O}_1(P_1) \mathcal{O}_2(P_2) \mathcal{O}_3(P_3) \rangle_{+} = \frac{g}{2} \pi^{\frac{d}{2}}\mathcal{E}^{+}_3 \Gamma\left( \frac{\displaystyle \sum_{i=1}^3 \Delta^{+}_n- d }{2}\right) \prod_{i <j}^3 \Gamma({\delta^{+}_{ij}}) (-P_{ij})^{-\delta^{+}_{ij}}
\end{equation}
\begin{equation}
\langle \mathcal{O}_1(P_1) \mathcal{O}_2(P_2) \mathcal{O}_3(P_3) \rangle_{-} = \frac{g}{2} \pi^{\frac{d}{2}}\mathcal{E}^{-}_3 \Gamma\left( \frac{\displaystyle \sum_{i=1}^3 \Delta^{-}_n- d }{2}\right) \prod_{i <j}^3 \Gamma({\delta^{-}_{ij}}) (-P_{ij})^{-\delta^{-}_{ij}}
\end{equation}
Consequently, the total scalar three point function can be finally expressed as:
\begin{eqnarray}
\langle \mathcal{O}_1(P_1) \mathcal{O}_2(P_2) \mathcal{O}_3(P_3) \rangle &=&  \frac{g}{2} \pi^{\frac{d}{2}}\Biggl\{\mathcal{E}^{+}_3 \Gamma\left( \frac{\displaystyle \sum_{i=1}^3 \Delta^{+}_n- d }{2}\right) \prod_{i <j}^3 \Gamma({\delta^{+}_{ij}}) (-P_{ij})^{-\delta^{+}_{ij}}\nonumber\\
&&~~~~~~~~~~~~~~~+\mathcal{E}^{-}_3 \Gamma\left( \frac{\displaystyle \sum_{i=1}^3 \Delta^{-}_n- d }{2}\right) \prod_{i <j}^3 \Gamma({\delta^{-}_{ij}}) (-P_{ij})^{-\delta^{-}_{ij}}\Biggr\}.~~~~~~~
\end{eqnarray}
In this specific example of dS space time, the three point {\it Mellin Barnes amplitudes} corrsponding to the time-ordered and anti-time-ordered contributions in the scalar three point functions are given by the following expressions:
\bea {\cal M}^{+}_3:&=&g~\Gamma\left( \frac{\displaystyle\sum_{n=1}^3 \Delta^{+}_n- d }{2}\right)=g~\Gamma\left( \displaystyle\frac{d}{4}\left[1+i\sum_{n=1}^3 \sqrt{\left(\frac{2m_nl}{d}\right)^2-1}\right]\right),\\
{\cal M}^{-}_3:&=&g~\Gamma\left( \frac{\displaystyle\sum_{n=1}^3 \Delta^{-}_n- d }{2}\right)=g~\Gamma\left( \displaystyle\frac{d}{4}\left[1-i\sum_{n=1}^3 \sqrt{\left(\frac{2m_nl}{d}\right)^2-1}\right]\right).\eea  
One can further consider few special cases further, to study the various outcomes of the three point function and the related {\it Mellin Barnes amplitude} derived in this section for the dS space time. First of all one can consider computing the three point function and the related {\it Mellin Barnes amplitude} from three identical scalar fields having the same mass parameter, $m$. In that case the expression the conformal dimension of each individual operators participating in the three point function computation will be further simplified by replacing $m_i$ with $m$. This further implies that the conformal dimension of the there operators are identical in this particular case and given by:
\bea \Delta^{\pm}:=\Delta^{\pm}_1=\Delta^{\pm}_2=\Delta^{\pm}_3=\frac{d}{2}\pm i\nu,~~~{\rm where}~~\nu=\sqrt{(ml)^2-\left(\frac{d}{2}\right)^2}.\eea
In this particular case, we further have the following simplifications:
\begin{equation}
\delta^{\pm}:=\delta^{\pm}_{12} = \delta^{\pm}_{23} =\delta^{\pm}_{13} = \frac{\Delta^{\pm}}{2}.
\end{equation}
In this case, the time ordered and the anti-time ordered contribution in the scalar three point function can be further simplified as:
\begin{eqnarray}
\langle \mathcal{O}_1(P_1) \mathcal{O}_2(P_2) \mathcal{O}_3(P_3) \rangle_{+} &=&\frac{g}{2} \pi^{\frac{d}{2}}\mathcal{E}^{+}_3 \Gamma\left( \frac{\displaystyle 3 }{2}\left(\frac{d}{2}+i\nu\right)-\frac{d}{2}\right) \Gamma\left(\frac{1}{2}\left(\frac{d}{2}+i\nu\right)\right)~\nonumber\\
&&~~~~~~~~~~~~~~~~~~~~~~~~~~~~~~\prod_{i <j}^3  (-P_{ij})^{-\frac{1}{2}\left(\frac{d}{2}+i\nu\right)},~~~~~~~~
\\
\langle \mathcal{O}_1(P_1) \mathcal{O}_2(P_2) \mathcal{O}_3(P_3) \rangle_{-} &=& \frac{g}{2} \pi^{\frac{d}{2}}\mathcal{E}^{-}_3 \Gamma\left(\frac{3}{2}\left(\frac{d}{2}-i\nu\right)-\frac{d}{2}\right) \Gamma\left(\frac{1}{2}\left(\frac{d}{2}-i\nu\right)\right)\nonumber\\
&&~~~~~~~~~~~~~~~~~~~~~~~~~~~~~~\prod_{i <j}^3  (-P_{ij})^{-\frac{1}{2}\left(\frac{d}{2}-i\nu\right)},~~~~~~~~
\end{eqnarray}
and the total scalar three point function in that case can be expressed 
as:
\begin{eqnarray}
\langle \mathcal{O}_1(P_1) \mathcal{O}_2(P_2) \mathcal{O}_3(P_3) \rangle &=& \frac{g}{2} \pi^{\frac{d}{2}}\Biggl[\mathcal{E}^{+}_3 \Gamma\left( \frac{\displaystyle 3 }{2}\left(\frac{d}{2}+i\nu\right)-\frac{d}{2}\right) \Gamma\left(\frac{1}{2}\left(\frac{d}{2}+i\nu\right)\right)~\prod_{i <j}^3  (-P_{ij})^{-\frac{1}{2}\left(\frac{d}{2}+i\nu\right)}\nonumber\\
&&~~~~~+\mathcal{E}^{-}_3 \Gamma\left(\frac{3}{2}\left(\frac{d}{2}-i\nu\right)-\frac{d}{2}\right) \Gamma\left(\frac{1}{2}\left(\frac{d}{2}-i\nu\right)\right)~\prod_{i <j}^3  (-P_{ij})^{-\frac{1}{2}\left(\frac{d}{2}-i\nu\right)}\Biggr],~~~~~~~~
\end{eqnarray}
and the corresponding {\it Mellin Barnes amplitudes} corrsponding to the time-ordered and anti-time-ordered contributions in the scalar three point functions are given by the following expressions:
\bea {\cal M}^{+}_3:=g~\Gamma\left( \frac{\displaystyle 3\Delta^{+}- d }{2}\right) &=&g~\Gamma\left( \displaystyle\frac{d}{4}\left[1+3i\sqrt{\left(\frac{2ml}{d}\right)^2-1}\right]\right)\nonumber\\
&=&g~\Gamma\left(\frac{3}{2}\left(\frac{d}{2}+i\nu\right)-\frac{d}{2}\right),\eea
\bea
{\cal M}^{-}_3:=g~ \Gamma\left( \frac{\displaystyle 3\Delta^{-}- d }{2}\right)&=&g~\Gamma\left( \displaystyle\frac{d}{4}\left[1-3i \sqrt{\left(\frac{2ml}{d}\right)^2-1}\right]\right)\nonumber\\
&=&g~\Gamma\left(\frac{3}{2}\left(\frac{d}{2}-i\nu\right)-\frac{d}{2}\right).\eea  
In this context, the factors $\mathcal{E}^{+}_3$ and $\mathcal{E}^{-}_3$ are given by the following expressions:
\be 
\mathcal{E}^{\pm}_3= \prod_{n=1}^3 \frac{ 1}{2 \pi^{\frac{d}{2}} \Gamma \left(1\pm i\nu \right)}= \frac{ 1}{8 \pi^{\frac{3d}{2}} \Gamma ^3\left(1\pm i\nu \right)}.\ee
Particularly in $d=3$ the conformal dimension of the identical operators can be written as:
\bea \Delta^{\pm}=\frac{3}{2}\left(1\pm i\sqrt{\left(\frac{2ml}{3}\right)^2-1}\right)=\frac{3}{2}\pm i\nu~~~{\rm with}~~\nu=\sqrt{(ml)^2-\frac{9}{4}}.\eea
 Here one can further check that for $d=3$ we get:
\begin{eqnarray}
&&\Delta^{+}=\frac{3}{2}+ i\nu~~~\longrightarrow~~~{\cal M}_3=g~\Gamma\left(\frac{3}{2}\left(\frac{3}{2}+ i\nu\right)-\frac{3}{2}\right),~~~~\nonumber\\
&&\langle \mathcal{O}_1(P_1) \mathcal{O}_2(P_2) \mathcal{O}_3(P_3) \rangle_{+} =\frac{g}{2}\pi^{\frac{3}{2}}\mathcal{E}^{+}_3\Gamma\left(\frac{3}{2}\left(\frac{3}{2}+ i\nu\right)-\frac{3}{2}\right)\Gamma\left(\frac{1}{2}\left(\frac{3}{2}+ i\nu\right)\right)\prod_{i <j}^3  (-P_{ij})^{-\frac{1}{2}\left(\frac{3}{2}+ i\nu\right)},~~~~~~~~~ \end{eqnarray}
\begin{eqnarray}
&&\Delta^{-}=\frac{3}{2}- i\nu~~~\longrightarrow~~~{\cal M}_3=g~\Gamma\left(\frac{3}{2}\left(\frac{3}{2}- i\nu\right)-\frac{3}{2}\right),~~~~\nonumber\\
&&\langle \mathcal{O}_1(P_1) \mathcal{O}_2(P_2) \mathcal{O}_3(P_3) \rangle_{-} =\frac{g}{2}\pi^{\frac{3}{2}}\mathcal{E}^{-}_3\Gamma\left(\frac{3}{2}\left(\frac{3}{2}- i\nu\right)-\frac{3}{2}\right)\Gamma\left(\frac{1}{2}\left(\frac{3}{2}- i\nu\right)\right)\prod_{i <j}^3  (-P_{ij})^{-\frac{1}{2}\left(\frac{3}{2}- i\nu\right)}.~~~~~~~~~~
\end{eqnarray}
where, the factors $\mathcal{E}^{+}_3$ and $\mathcal{E}^{-}_3$ are given by the following expressions for $d=3$ case:
\be 
\mathcal{E}^{\pm}_3= \prod_{n=1}^3 \frac{ 1}{2 \pi^{\frac{3}{2}} \Gamma \left(1\pm i\nu \right)}= \frac{ 1}{8 \pi^{\frac{9}{2}} \Gamma ^3\left(1\pm i\nu \right)}.\ee
Now in $d=3$ for massless scalar fields the conformal dimension of the operators can be written as:
\bea \Delta^{+}=0, ~~\Delta^{-}=3.\eea
 Here one can further check that for $d=3$ we get:
\begin{eqnarray}
&&\Delta^{+}=0~~~\longrightarrow~~~{\cal M}_3=g~\Gamma\left(-\frac{3}{2}\right)=\frac{4g\sqrt{\pi}}{3},~~~~~~~~\langle \mathcal{O}_1(P_1) \mathcal{O}_2(P_2) \mathcal{O}_3(P_3) \rangle_{+} \rightarrow \infty, \\
&&\Delta^{-}=3~~~\longrightarrow~~~{\cal M}_3=g~\Gamma\left(3\right)=2g,~~~~\langle \mathcal{O}_1(P_1) \mathcal{O}_2(P_2) \mathcal{O}_3(P_3) \rangle_{-} =\frac{g}{2}\pi^{2}\mathcal{E}^{-}_3\prod_{i <j}^3  (-P_{ij})^{-\frac{3}{2}}.~~~~~~~~~~
\end{eqnarray}
where, the factor $\mathcal{E}^{-}_3$ is given by for the massless scalar field by the following expression for $d=3$ case:
\be 
\mathcal{E}^{-}_3= \prod_{n=1}^3 \frac{ 1}{2 \pi^{\frac{3}{2}} \Gamma \left(\frac{5}{2} \right)}= \prod_{n=1}^3 \frac{ 2}{3 \pi^{2} }= \frac{ 8}{27\pi^6}.\ee
The above mentioned pathology is not the outcome of only $d=3$, but will valid for any arbitrary spatial dimension with massless scalar fields. This can be demonstrated for any arbitrary $d$ with massless scalar fields as following:
\bea \Delta^{+}=0, ~~\Delta^{-}=d.\eea
Consequently, we get the following expressions for the {\it Mellin Barnes amplitude} and the three point function:
\begin{eqnarray}
&&\Delta=0~~~\longrightarrow~~~{\cal M}_3=g~\Gamma\left(-\frac{d}{2}\right),~~~~\langle \mathcal{O}_1(P_1) \mathcal{O}_2(P_2) \mathcal{O}_3(P_3) \rangle \rightarrow \infty, \\
&&\Delta=d~~~\longrightarrow~~~{\cal M}_3=g~\Gamma\left(d\right),~~\langle \mathcal{O}_1(P_1) \mathcal{O}_2(P_2) \mathcal{O}_3(P_3) \rangle = \frac{g}{2}\pi^{d/2}\Gamma(d)\Gamma\left(\frac{d}{2}\right)\mathcal{E}^{-}_3~\prod_{i <j}^3  P_{ij}^{-\frac{d}{2}},~~~~~~~~~~
\end{eqnarray}
 where, the factor $\mathcal{E}^{-}_3$ is given by for the massless scalar field by the following expression:
\be 
\mathcal{E}^{-}_3= \prod_{n=1}^3 \frac{ 1}{2 \pi^{\frac{d}{2}} \Gamma \left(1+\frac{d}{2} \right)}= \frac{ 1}{8 \pi^{\frac{3d}{2}}  \Gamma^3 \left(1+\frac{d}{2} \right)}.\ee
\section{Four point function}
\subsection{From the scalar field in AdS space}
\label{sec:4ads}

In this section our prime objective is to look back to the computation of the four point function for scalar fields in the background of Anti-de Sitter space time in presence of scalar exchange interaction.  In this context,  let us consider the $s$-channel diagram.  In this case one can consider two three point interaction appearing at the point $X_1$ and $X_2$ over which we have to integrate at the end.  Here the corresponding amplitude can be describe by the following four-point function:
\bea \langle {\cal O}_1(P_1){\cal O}_2(P_2){\cal O}_3(P_3){\cal O}_4(P_4)\rangle &=& g^2 
\int^{\infty}_{0}dX_1~\int^{\infty}_{0}dX_2~K_{\Delta_1}(X_1,P_1)K_{\Delta_2}(X_1,P_2)~\nonumber\\
&&~~~~~~~~~~~~~~~~~~~~~~~~{\cal G}(X_1,X_2)~K_{\Delta_3}(X_2,P_3)K_{\Delta_4}(X_2,P_4). ~~~~~~~~~~~~\eea
Here it is important to note that the bulk-to-bulk propagator ${\cal G}(X_1,X_2)$ can be written as:
\bea {\cal G}(X_1,X_2)&=&\int^{+i\infty}_{-i\infty}\frac{dc}{2\pi i}~{\cal F}_{\delta,0}(c)~\int_{\partial{\rm AdS}}~dQ~\int~\widetilde{d^2s_c}~\exp(2(sQ.X_1+\bar{s}Q.X_2)),\eea
where ${\cal F}_{\delta,0}(c)$ and $\widetilde{d^2s_c}$ is defined as:
\bea {\cal F}_{\delta,0}(c)&=&\frac{1}{2\pi^{2h}\Gamma(c)\Gamma(-c)\left\{\left(\delta-h\right)^2-c^2\right\}}~~~~~~~~~{\rm where}~~~h=\frac{d}{2},~\delta=\frac{\Delta}{2}~~~~\\
\widetilde{d^2s_c}&=& \frac{ds}{s}~\frac{d\bar{s}}{\bar{s}}~s^{h+c}~\bar{s}^{h-c}.\eea
Further substituting the above mentioned expression for the bulk-to-bulk propagator ${\cal G}(X_1,X_2)$ in the above mentioned expression for the four-point $s$-channel contribution one can write:
\bea 
\langle {\cal O}_1(P_1){\cal O}_2(P_2){\cal O}_3(P_3){\cal O}_4(P_4)\rangle &=&  \int^{+i\infty}_{-i\infty}\frac{dc}{2\pi i}~{\cal F}_{\delta,0}(c)~\nonumber\\
&&~~~~~~\int_{\partial{\rm AdS}}~dQ~{\cal A}_{h+c,\Delta_1,\Delta_2}(Q_{+},P_1,P_2)~{\cal A}_{h-c,\Delta_3,\Delta_4}(Q_{-},P_3,P_4)
\nonumber\\
&&\eea
where the functions ${\cal A}_{h+c,\Delta_1,\Delta_2}(Q_{+},P_1,P_2)$ and ${\cal A}_{h-c,\Delta_3,\Delta_4}(Q_{-},P_3,P_4)$ are defined by the following expressions:
\bea {\cal A}_{h+c,\Delta_1,\Delta_2}(Q_{+},P_1,P_2):&=& g\int^{\infty}_{0}\frac{dt_1}{t_1}~\frac{dt_2}{t_2}~\frac{ds}{s}~t^{\Delta_1}_1~t^{\Delta_2}_2~s^{h+c}~\nonumber\\
&&~~~~~~~\int_{{\rm AdS}}~dX_1~\exp(2(t_1P_1+t_2P_2+sQ).X_1)~,~~~~~~~~~\eea
and 
\bea {\cal A}_{h-c,\Delta_3,\Delta_4}(Q_{-},P_3,P_4):&=& g\int^{\infty}_{0}\frac{dt_3}{t_3}~\frac{dt_4}{t_4}~\frac{d\bar{s}}{\bar{s}}~t^{\Delta_1}_3~t^{\Delta_2}_4~\bar{s}^{h-c}~\nonumber\\
&&~~~~~~~\int_{{\rm AdS}}~dX_2~\exp(2(t_3P_3+t_4P_4+\bar{s}Q).X_2)~.~~~~~~~~~\eea
Now if we closely look into the above mentioned two amplitudes then we see that both of them are representing three point amplitudes which we have explicitly evaluated in the previous sections in detail.  

Now,  since all the bulk-to-bulk propagators factorise in the above mentioned specific way,  any $n$-point scattering amplitude can be expressed in terms of the three point amplitudes which are connected with each other.  Here for this computation we have adopt the following notation:
\bea &&{\cal A}_{h\pm c_k,\Delta_i,\Delta_j}(Q_{\pm},P_i,P_j):\equiv {\cal A}(c^{\pm}_k,i,j)~~~~~~{\rm where}~~~k={\rm Number~of~three-point~amplitudes}.\nonumber\\
&& \eea
Now,  to compute the mentioned four-point amplitude in the context of AdS space the usual trick is to introduce Schwinger parameters,  $t$ and $s$ which suppose to appear in the exponential part of the amplitude integral.  Here it important to note that,  these parameters also appearing in the expression for the bulk-to-bulk propagators explicitly have written in the previous page.  The simplest way to deal with these amplitude integrals is to first of perform the the integration over $X$ variables.  For example,  in the case of the computing the four -point function it would best if one can perform the integrals over the $X_1$ and $X_2$ variables.  After performing this job we obtain the following simplified compact result:
\bea {\cal A}(c^{\pm},i,j)&=&g~\pi^h~\Gamma\left(\frac{\Delta_i+\Delta_j+(h\pm c)-2h}{2}\right)~\nonumber\\
&&~~~~~~\times\int^{\infty}_{0}\frac{dt_i}{t_i}~\frac{dt_j}{t_j}~\frac{ds}{s}~t^{\Delta_i}_i~t^{\Delta_j}_j~s^{h\pm c}~\exp\left(-t_it_jP_{ij}+2sQ.(t_iP_i+t_jP_j)\right).~~~~~~~~~~~\eea
Next job is to perform the integral over the $Q$ variable for which we use the following result, which we have previously used in this paper:
\bea &&\int^{\infty}_{0}~\frac{ds}{s}~\frac{d\bar{s}}{\bar{s}}~s^{h+c}~\bar{s}^{h-c}~\int_{\partial{\rm AdS}}~dQ~\exp(2Q.(sP_i+\bar{s}P_j))\nonumber\\
&&~~~~~~~~~~~~~~~~~~=2\pi^h~ \int^{\infty}_{0}~\frac{ds}{s}~\frac{d\bar{s}}{\bar{s}}~s^{h+c}~\bar{s}^{h-c}~\exp((sP_i+\bar{s}P_j)^2).\eea
using this crucial integral identity we get the following simplified result for the four-point amplitude in the AdS space:
\bea 
&&\langle{\cal O}_1(P_1){\cal O}_2(P_2){\cal O}_3(P_3){\cal O}_4(P_4)\rangle \nonumber\\
&&=  \int^{+i\infty}_{-i\infty}\frac{dc}{2\pi i}~{\cal F}_{\delta,0}(c)~\int~\widetilde{d^2s_c}~\Gamma\left(\frac{\Delta_1+\Delta_2+c-h}{2}\right)\Gamma\left(\frac{\Delta_3+\Delta_4-c-h}{2}\right)\nonumber\\
&&~~~~\times\int^{\infty}_{0}~\frac{dt_1}{t_1}\frac{dt_2}{t_2}\frac{dt_3}{t_3}\frac{dt_4}{t_4}~t^{\Delta_1}_1 t^{\Delta_2}_2 t^{\Delta_3}_3 t^{\Delta_4}_4~\nonumber\\
&&~~~~~~~~\times\exp\left(-(1+s^2)t_1t_2P_{12}-(1+\bar{s}^2)t_3t_4P_{34}-s\bar{s}\left(t_1t_3P_{13}+t_1t_4P_{14}+t_2t_3P_{23}+t_2t_4P_{24}\right)\right).\nonumber\\
&&\eea  
Next, our job is to evaluate the above amplitude integral explicitly.  Here we use the well known Symanzik's star formula to evaluate the following Mellin-Barnes amplitude integral appearing in the above mentioned expression for the four-point amplitude:
\bea {\cal M}_4=2~\int^{i\infty}_{-i\infty}\frac{dc}{2\pi i}~{\cal F}_{\delta,0}(c)~{\cal I}_{\bf Bulk}(12,h,c)~{\cal I}_{\bf Bulk}(34,h,-c),\eea
where,  the two integrands ${\cal I}_{\bf Bulk}(12,h,c)$ and ${\cal I}_{\bf Bulk}(34,h,-c)$ are explicitly written as:
\bea {\cal I}_{\bf Bulk}(12,h,c)&=&g~\pi^h~\Gamma\left(\frac{\Delta_1+\Delta_2+c-h}{2}\right)~\int^{\infty}_{0}~\frac{ds}{s}~~s^{{\displaystyle h+c-\sum_{(ij)}}^{'}\displaystyle \delta_{ij}}~(1+s^2)^{-\delta_{12}},~~~~~~~~~~\\
{\cal I}_{\bf Bulk}(34,h,-c)&=&g~\pi^h~\Gamma\left(\frac{\Delta_3+\Delta_4-c-h}{2}\right)~\int^{\infty}_{0}~\frac{d\bar{s}}{\bar{s}}~~\bar{s}^{{\displaystyle h-c-\sum_{(ij)}}^{'}\displaystyle \delta_{ij}}~(1+\bar{s}^2)^{-\delta_{34}}.\eea
Now,  one can simplify the expression for the Mellin-Barnes amplitude integral in terms of the Mandelstam variables:
\bea {\cal M}_4(s_{12})=\frac{g^2}{
\displaystyle \Gamma\left(\frac{\Delta_1+\Delta_2-s_{12}}{2}\right)\Gamma\left(\frac{\Delta_3+\Delta_4-s_{12}}{2}\right)}~ \int^{+i\infty}_{-i\infty}\frac{dc}{2\pi i}~\frac{{\cal R}_h(c){\cal R}_h(-c)}{\left\{\left(\delta-h\right)^2-c^2\right\}},~~~~~~\eea
where we define the functions ${\cal R}_h(c)$ and ${\cal R}_h(-c)$ by the following expressions:
\bea {\cal R}_h(c):&=&\frac{1}{2\Gamma(c)}~\Gamma\left(\frac{\Delta_1+\Delta_2+c-h}{2}\right)\Gamma\left(\frac{\Delta_3+\Delta_4+c-h}{2}\right),\\
{\cal R}_h(-c):&=&\frac{1}{2\Gamma(-c)}~\Gamma\left(\frac{\Delta_1+\Delta_2-c-h}{2}\right)\Gamma\left(\frac{\Delta_3+\Delta_4-c-h}{2}\right).\eea
Finally,  the above mentioned  Mellin-Barnes amplitude integral can be expressed after performing the integral in complex plane as:
\bea {\cal M}_4(s_{12})&=&\frac{g^2}{2}~\frac{1}{(s_{12}-\delta)}~\frac{1}{\Gamma(1+\delta-h)}~\Gamma\left(\frac{\Delta_1+\Delta_2+\delta-h}{2}\right)\Gamma\left(\frac{\Delta_3+\Delta_4+\delta-h}{2}\right)\nonumber\\
&&\times {}_3F_1\left(\frac{2+\delta-\Delta_1-\Delta_2}{2},\frac{2+\delta-\Delta_3-\Delta_4}{2},\frac{\delta-s_{12}}{2},\frac{2+\delta-s_{12}}{2},1+\delta-h,1\right).\nonumber\\
&=&\sum^{\infty}_{n=0}\frac{P^{\delta}_n}{s_{12}-\delta-2n}{\cal V}^{\Delta_1,\Delta_2,\delta}_{[0,0,n]}{\cal V}^{\Delta_3,\Delta_4,\delta}_{[0,0,n]},\eea
where the three point vertices and the normalised propagators are defined as:
\bea {\cal V}^{\Delta_1,\Delta_2,\delta}_{[0,0,0]}&=&g~\Gamma\left(\frac{\Delta_1+\Delta_2+\delta-2h}{2}\right),\\
{\cal V}^{\Delta_3,\Delta_4,\delta}_{[0,0,0]}&=&g~\Gamma\left(\frac{\Delta_3+\Delta_4+\delta-2h}{2}\right),\\
{\cal V}^{\Delta_1,\Delta_2,\delta}_{[0,0,n]}&=&{\cal V}^{\Delta_1,\Delta_2,\delta}_{[0,0,0]}~\frac{\displaystyle \Gamma\left(1-\frac{1}{2}\left(\Delta_1+\Delta_2-\delta\right)+n\right)}{\displaystyle \Gamma\left(1-\frac{1}{2}\left(\Delta_1+\Delta_2-\delta\right)\right)}\nonumber\\
&=&g~\Gamma\left(\frac{\Delta_1+\Delta_2+\delta-2h}{2}\right)~\frac{\displaystyle \Gamma\left(1-\frac{1}{2}\left(\Delta_1+\Delta_2-\delta\right)+n\right)}{\displaystyle \Gamma\left(1-\frac{1}{2}\left(\Delta_1+\Delta_2-\delta\right)\right)},\\
{\cal V}^{\Delta_3,\Delta_4,\delta}_{[0,0,n]}&=&{\cal V}^{\Delta_3,\Delta_4,\delta}_{[0,0,0]}~\frac{\displaystyle \Gamma\left(1-\frac{1}{2}\left(\Delta_3+\Delta_4-\delta\right)+n\right)}{\displaystyle \Gamma\left(1-\frac{1}{2}\left(\Delta_3+\Delta_4-\delta\right)\right)}\nonumber\\
&=&g~\Gamma\left(\frac{\Delta_3+\Delta_4+\delta-2h}{2}\right)~\frac{\displaystyle \Gamma\left(1-\frac{1}{2}\left(\Delta_3+\Delta_4-\delta\right)+n\right)}{\displaystyle \Gamma\left(1-\frac{1}{2}\left(\Delta_3+\Delta_4-\delta\right)\right)}.
\eea
Also the normalization factor $P^{\Delta}_n$ is defined as:
\bea P^{\Delta}_n&=& \frac{1}{2n! ~\Gamma(1+\delta-h+n)}~~~~~~~~~{\rm where}~~~h=\frac{d}{2},~\delta=\frac{\Delta}{2}.~~~~\eea
Here at the end the Mellin amplitude is expressed as an infinite sum of products of three point vertices and propagators.  It is important to note that the sum runs over the propagating fields,  which include a field with conformal dimension $\delta$ and its descendants with dimension $\delta+2n$.

Now here it is important to note that,  in this result a set of Feynman rules for Mellin amplitudes are appearing which are appended below point-wise:
\begin{enumerate}
\item \textbf{\underline{Rule~I:}}\\
In the each internal line of the bulk Witten diagram an infinite sum of propagating fields are associated out of which one is identified to be the primary field and rest of the infinite possibilities are descendent fields. 

\item  \textbf{\underline{Rule~II:}}\\
In each vertex one can immediate associate the vertex factor ${\cal V}^{\Delta_1,\Delta_2,\Delta_3}_{[l,m,n]}$.

\item  \textbf{\underline{Rule~III:}}\\
For the $i$-th external line a normalization factor $\frac{\displaystyle 1}{\displaystyle\Gamma\left(1+\Delta_i+n-h\right)}$ where $h=\displaystyle \frac{d}{2}$.

\end{enumerate} 

Now once we compute the higher point (five point and six point) amplitudes then one can easily cross check the usefulness of the above mentioned Feynman rules for the bulk Witten diagrams.  
\subsection{From the scalar field in dS space} 
\label{sec:4ds}

In this section our prime objective is to look back to the computation of the four point function for scalar fields in the background of de Sitter space time in presence of scalar exchange interaction.  In this context,  let us consider the $s$-channel diagram in de Sitter space.  In this case one can consider two three point interaction appearing at the point $X_1$ and $X_2$ over which we have to integrate at the end.  Here the corresponding amplitude can be describe by the following four-point function:
\bea \langle {\cal O}_1(P_1){\cal O}_2(P_2){\cal O}_3(P_3){\cal O}_4(P_4)\rangle &=& \langle {\cal O}_1(P_1){\cal O}_2(P_2){\cal O}_3(P_3){\cal O}_4(P_4)\rangle_{+}\nonumber\\
&&+\langle {\cal O}_1(P_1){\cal O}_2(P_2){\cal O}_3(P_3){\cal O}_4(P_4)\rangle_{-},\eea
where each of the individual contributions appearing in the above mentioned expression are appended below:
\bea \langle {\cal O}_1(P_1){\cal O}_2(P_2){\cal O}_3(P_3){\cal O}_4(P_4)\rangle_{+} &=& g^2 
\int^{\infty}_{0}dX_1~\int^{\infty}_{0}dX_2~K_{\Delta^{+}_1}(\mp iX_1,P_1)K_{\Delta^{+}_2}(\mp iX_1,P_2)~\nonumber\\
&&~~~~~~~{\cal G}_{+}(\mp iX_1,\mp iX_2)~K_{\Delta^{+}_3}(\mp iX_2,P_3)K_{\Delta^{+}_4}(\mp iX_2,P_4). ~~~~~~~~~~~~\\
 \langle {\cal O}_1(P_1){\cal O}_2(P_2){\cal O}_3(P_3){\cal O}_4(P_4)\rangle_{-} &=& g^2 
\int^{\infty}_{0}dX_1~\int^{\infty}_{0}dX_2~K_{\Delta^{-}_1}(\pm iX_1,P_1)K_{\Delta^{-}_2}(\pm iX_1,P_2)~\nonumber\\
&&~~~~~~~{\cal G}_{-}(\pm iX_1,\pm iX_2)~K_{\Delta^{-}_3}(\pm iX_2,P_3)K_{\Delta^{-}_4}(\pm iX_2,P_4). ~~~~~~~~~~~~\eea
In this context, bulk to boundary propagators in dS space after analytically continuing from AdS space we get:
\bea {\cal K}_{\Delta^{+}_n}\left(\mp i X_1,P\right) :&=&\frac{\Gamma\left(\Delta^{+}_n-\frac{d}{2}+1\right)}{\sqrt{\pi}}~K_{\Delta^{+}_n}\left(\mp i X_1,P\right)=\frac{\mathcal{C}_{\Delta^{+}_n} }{(-2 P \cdot (\mp i X_1))^{\Delta^{+}_n}},~~\forall~n=1,2,3,4,\\
{\cal K}_{\Delta^{-}_n}\left(\pm i X_2,P\right) :&=&\frac{\Gamma\left(\Delta^{-}_n-\frac{d}{2}+1\right)}{\sqrt{\pi}}~K_{\Delta^{-}_n}\left(\pm i X_2,P\right)=\frac{\mathcal{C}_{\Delta^{-}_n} }{(-2 P \cdot (\mp i X_1))^{\Delta^{-}_n}},~~\forall~n=1,2,3,4,~~~~~~~~~\eea
where we define $C_{\Delta^{\pm}_n}~~\forall~~n=1,2,3,4$ by the following expression:
\bea
\mathcal{C}_{\Delta^{\pm}_n} &=& \frac{ \Gamma({\Delta^{\pm}_n})}{2 \pi^{\frac{d}{2}} \Gamma \left({\Delta^{\pm}_n}-\frac{d}{2}+1 \right)}\,.
\eea
Here it is important to note that the bulk-to-bulk propagators ${\cal G}_{+}(\mp iX_1,\mp iX_2)$ and ${\cal G}_{-}(\mp iX_1,\mp iX_2)$ can be written in the present context as: 
\bea {\cal G}_{+}(\mp iX_1,\mp iX_2)&=&\int^{+i\infty}_{-i\infty}\frac{dc}{2\pi i}~{\cal F}_{\delta^{+},0}(c)~\int_{\partial{\rm dS}}~dQ~\int~\widetilde{d^2s_c}~\exp(\mp 2i(sQ.X_1+\bar{s}Q.X_2)),~~~~~~~~~~~~~~\\
{\cal G}_{-}(\mp iX_1,\mp iX_2)&=&\int^{+i\infty}_{-i\infty}\frac{dc}{2\pi i}~{\cal F}_{\delta^{-},0}(c)~\int_{\partial{\rm dS}}~dQ~\int~\widetilde{d^2s_c}~\exp(\pm 2i(sQ.X_1+\bar{s}Q.X_2)).\eea
where ${\cal F}_{\delta^{+},0}(c)$, ${\cal F}_{\delta^{-},0}(c)$ and $\widetilde{d^2s_c}$ is defined as:
\bea {\cal F}_{\delta^{+},0}(c)&=&\frac{1}{2\pi^{2h}\Gamma(c)\Gamma(-c)\left\{\left(\delta^{+}-h\right)^2-c^2\right\}}~~~~~~~~~{\rm where}~~~h=\frac{d}{2},~\delta^{+}=\frac{\Delta^{+}}{2}~~~~\\
{\cal F}_{\delta^{-},0}(c)&=&\frac{1}{2\pi^{2h}\Gamma(c)\Gamma(-c)\left\{\left(\delta^{-}-h\right)^2-c^2\right\}}~~~~~~~~~{\rm where}~~~h=\frac{d}{2},~\delta^{-}=\frac{\Delta^{-}}{2}~~~~\\
\widetilde{d^2s_c}&=& \frac{ds}{s}~\frac{d\bar{s}}{\bar{s}}~s^{h+c}~\bar{s}^{h-c}.\eea
Further substituting the above mentioned expression for thebulk-to-bulk propagators ${\cal G}_{+}(\mp iX_1,\mp iX_2)$ and ${\cal G}_{-}(\mp iX_1,\mp iX_2)$ in the previously mentioned expression for the four-point $s$-channel contributions one can write:
\bea 
\langle {\cal O}_1(P_1){\cal O}_2(P_2){\cal O}_3(P_3){\cal O}_4(P_4)\rangle_{+} &=&  \int^{+i\infty}_{-i\infty}\frac{dc}{2\pi i}~{\cal F}_{\delta^{+},0}(c)~\nonumber\\
&&~~~~~~\int_{\partial{\rm dS}}~dQ~{\cal A}_{h+c,\Delta^{+}_1,\Delta^{+}_2}(Q_{+},P_1,P_2)~{\cal A}_{h-c,\Delta^{+}_3,\Delta^{+}_4}(Q_{-},P_3,P_4)
\nonumber\\
&&\\
\langle {\cal O}_1(P_1){\cal O}_2(P_2){\cal O}_3(P_3){\cal O}_4(P_4)\rangle_{-} &=&  \int^{+i\infty}_{-i\infty}\frac{dc}{2\pi i}~{\cal F}_{\delta^{-},0}(c)~\nonumber\\
&&~~~~~~\int_{\partial{\rm dS}}~dQ~{\cal A}_{h+c,\Delta^{-}_1,\Delta^{-}_2}(Q_{+},P_1,P_2)~{\cal A}_{h-c,\Delta^{-}_3,\Delta^{-}_4}(Q_{-},P_3,P_4)
\nonumber\\
&&\eea
where the functions ${\cal A}_{h+c,\Delta^{+}_1,\Delta^{+}_2}(Q_{+},P_1,P_2)$,  ${\cal A}_{h-c,\Delta^{+}_3,\Delta^{+}_4}(Q_{-},P_3,P_4)$,  ${\cal A}_{h+c,\Delta^{-}_1,\Delta^{-}_2}(Q_{+},P_1,P_2)$,  ${\cal A}_{h-c,\Delta^{-}_3,\Delta^{-}_4}(Q_{-},P_3,P_4)$ are defined by the following expressions:
\bea {\cal A}_{h+c,\Delta^{+}_1,\Delta^{+}_2}(Q_{+},P_1,P_2):&=& g\int^{\infty}_{0}\frac{dt_1}{t_1}~\frac{dt_2}{t_2}~\frac{ds}{s}~t^{\Delta^{+}_1}_1~t^{\Delta^{+}_2}_2~s^{h+c}~\nonumber\\
&&~~~~~~~\int_{{\rm dS}}~dX_1~\exp(\mp 2i(t_1P_1+t_2P_2+sQ).X_1)~,~~~~~~~~~\\
{\cal A}_{h+c,\Delta^{-}_1,\Delta^{-}_2}(Q_{+},P_1,P_2):&=& g\int^{\infty}_{0}\frac{dt_1}{t_1}~\frac{dt_2}{t_2}~\frac{ds}{s}~t^{\Delta^{-}_1}_1~t^{\Delta^{-}_2}_2~s^{h+c}~\nonumber\\
&&~~~~~~~\int_{{\rm dS}}~dX_1~\exp(\pm 2i(t_1P_1+t_2P_2+sQ).X_1)~,~~~~~~~~~\eea
and  
\bea {\cal A}_{h-c,\Delta^{+}_3,\Delta^{+}_4}(Q_{-},P_3,P_4):&=& g\int^{\infty}_{0}\frac{dt_3}{t_3}~\frac{dt_4}{t_4}~\frac{d\bar{s}}{\bar{s}}~t^{\Delta^{+}_1}_3~t^{\Delta^{+}_2}_4~\bar{s}^{h-c}~\nonumber\\
&&~~~~~~~\int_{{\rm dS}}~dX_2~\exp(\mp 2i(t_3P_3+t_4P_4+\bar{s}Q).X_2)~,~~~~~~~~~\\
{\cal A}_{h-c,\Delta^{-}_3,\Delta^{-}_4}(Q_{-},P_3,P_4):&=& g\int^{\infty}_{0}\frac{dt_3}{t_3}~\frac{dt_4}{t_4}~\frac{d\bar{s}}{\bar{s}}~t^{\Delta^{-}_1}_3~t^{\Delta^{-}_2}_4~\bar{s}^{h-c}~\nonumber\\
&&~~~~~~~\int_{{\rm dS}}~dX_2~\exp(\pm 2i(t_3P_3+t_4P_4+\bar{s}Q).X_2)~.~~~~~~~~~
\eea
Here for the further computational simplification we have adopt the following notation in de Sitter space:
\bea &&{\cal A}_{h\pm c_k,\Delta^{\pm}_i,\Delta^{\pm}_j}(Q_{\pm},P_i,P_j):\equiv {\cal A}(c^{\pm}_k,i^{\pm},j^{\pm})~~~~~~{\rm where}~~~k={\rm Number~of~three-point~amplitudes}.\nonumber\\
&& \eea
Now,  to compute the mentioned four-point amplitude in dS space the usual trick is to introduce Schwinger parameters,  $t$ and $s$ which suppose to appear in the exponential part of the amplitude integral.  After introducing this parametrization we obtain the following simplified compact result:
\bea {\cal A}(c^{\pm},i^{+},j^{+})&=&g~\pi^h~\Gamma\left(\frac{\Delta^{+}_i+\Delta^{+}_j+(h\pm c)-2h}{2}\right)~\nonumber\\
&&~~~~~~\times\int^{\infty}_{0}\frac{dt_i}{t_i}~\frac{dt_j}{t_j}~\frac{ds}{s}~t^{\Delta^{+}_i}_i~t^{\Delta^{+}_j}_j~s^{h\pm c}~\exp\left(-t_it_jP_{ij}\mp 2isQ.(t_iP_i+t_jP_j)\right),~~~~~~~~~~~\eea
and
\bea
{\cal A}(c^{\pm},i^{-},j^{-})&=&g~\pi^h~\Gamma\left(\frac{\Delta^{-}_i+\Delta^{-}_j+(h\pm c)-2h}{2}\right)~\nonumber\\
&&~~~~~~\times\int^{\infty}_{0}\frac{dt_i}{t_i}~\frac{dt_j}{t_j}~\frac{ds}{s}~t^{\Delta^{-}_i}_i~t^{\Delta^{-}_j}_j~s^{h\pm c}~\exp\left(-t_it_jP_{ij}\pm 2isQ.(t_iP_i+t_jP_j)\right).~~~~~~~~~~~\eea   
Next job is to perform the integral over the $Q$ variable which can be done as:
\bea &&\int^{\infty}_{0}~\frac{ds}{s}~\frac{d\bar{s}}{\bar{s}}~s^{h+c}~\bar{s}^{h-c}~\int_{\partial{\rm dS}}~dQ~\exp(\mp 2iQ.(sP_i+\bar{s}P_j))\nonumber\\
&&~~~~~~~~~~~~~~~~~~=2\pi^h~ \int^{\infty}_{0}~\frac{ds}{s}~\frac{d\bar{s}}{\bar{s}}~s^{h+c}~\bar{s}^{h-c}~\exp(\mp i(sP_i+\bar{s}P_j)^2),\\
&&\int^{\infty}_{0}~\frac{ds}{s}~\frac{d\bar{s}}{\bar{s}}~s^{h+c}~\bar{s}^{h-c}~\int_{\partial{\rm dS}}~dQ~\exp(\pm 2iQ.(sP_i+\bar{s}P_j))\nonumber\\
&&~~~~~~~~~~~~~~~~~~=2\pi^h~ \int^{\infty}_{0}~\frac{ds}{s}~\frac{d\bar{s}}{\bar{s}}~s^{h+c}~\bar{s}^{h-c}~\exp(\pm i (sP_i+\bar{s}P_j)^2).\eea
using this crucial integral identities we get the following simplified result for the four-point amplitude in the dS space: 
\bea 
&&\langle{\cal O}_1(P_1){\cal O}_2(P_2){\cal O}_3(P_3){\cal O}_4(P_4)\rangle_{+} \nonumber\\
&&=  \int^{+i\infty}_{-i\infty}\frac{dc}{2\pi i}~{\cal F}_{\delta^{+},0}(c)~\int~\widetilde{d^2s_c}~\Gamma\left(\frac{\Delta^{+}_1+\Delta^{+}_2+c-h}{2}\right)\Gamma\left(\frac{\Delta^{+}_3+\Delta^{+}_4-c-h}{2}\right)\nonumber\\
&&~~~~\times\int^{\infty}_{0}~\frac{dt_1}{t_1}\frac{dt_2}{t_2}\frac{dt_3}{t_3}\frac{dt_4}{t_4}~t^{\Delta^{+}_1}_1 t^{\Delta^{+}_2}_2 t^{\Delta^{+}_3}_3 t^{\Delta^{+}_4}_4~\nonumber\\
&&~~~~~~~~\times\exp\left(-(1+s^2)t_1t_2P_{12}-(1+\bar{s}^2)t_3t_4P_{34}-s\bar{s}\left(t_1t_3P_{13}+t_1t_4P_{14}+t_2t_3P_{23}+t_2t_4P_{24}\right)\right).\nonumber\\
&&\eea\bea
&&\langle{\cal O}_1(P_1){\cal O}_2(P_2){\cal O}_3(P_3){\cal O}_4(P_4)\rangle_{-} \nonumber\\
&&=  \int^{+i\infty}_{-i\infty}\frac{dc}{2\pi i}~{\cal F}_{\delta^{-},0}(c)~\int~\widetilde{d^2s_c}~\Gamma\left(\frac{\Delta^{-}_1+\Delta^{-}_2+c-h}{2}\right)\Gamma\left(\frac{\Delta^{-}_3+\Delta^{-}_4-c-h}{2}\right)\nonumber\\
&&~~~~\times\int^{\infty}_{0}~\frac{dt_1}{t_1}\frac{dt_2}{t_2}\frac{dt_3}{t_3}\frac{dt_4}{t_4}~t^{\Delta^{-}_1}_1 t^{\Delta^{-}_2}_2 t^{\Delta^{-}_3}_3 t^{\Delta^{-}_4}_4~\nonumber\\
&&~~~~~~~~\times\exp\left(-(1+s^2)t_1t_2P_{12}-(1+\bar{s}^2)t_3t_4P_{34}-s\bar{s}\left(t_1t_3P_{13}+t_1t_4P_{14}+t_2t_3P_{23}+t_2t_4P_{24}\right)\right).\nonumber\\
&&\eea
Further we use the well known Symanzik's star formula to evaluate the following Mellin-Barnes amplitude integrals appearing in the above mentioned expressions for the four-point amplitudes in de Sitter space:
\bea && {\cal M}^{+}_4=2~\int^{i\infty}_{-i\infty}\frac{dc}{2\pi i}~{\cal F}_{\delta^{+},0}(c)~{\cal I}^{+}_{\bf Bulk}(12,h,c)~{\cal I}^{+}_{\bf Bulk}(34,h,-c),\\
&& {\cal M}^{-}_4=2~\int^{i\infty}_{-i\infty}\frac{dc}{2\pi i}~{\cal F}_{\delta^{-},0}(c)~{\cal I}^{-}_{\bf Bulk}(12,h,c)~{\cal I}^{-}_{\bf Bulk}(34,h,-c),
\eea
where,  the four integrands ${\cal I}^{+}_{\bf Bulk}(12,h,c)$,  ${\cal I}^{-}_{\bf Bulk}(12,h,c)$,  ${\cal I}^{+}_{\bf Bulk}(34,h,-c)$ and ${\cal I}^{-}_{\bf Bulk}(34,h,-c)$ are explicitly written in de Sitter space as:
\bea {\cal I}^{+}_{\bf Bulk}(12,h,c)&=&g~\pi^h~\Gamma\left(\frac{\Delta^{+}_1+\Delta^{+}_2+c-h}{2}\right)~\int^{\infty}_{0}~\frac{ds}{s}~~s^{{\displaystyle h+c-\sum_{(ij)}}^{'}\displaystyle \delta^{+}_{ij}}~(1+s^2)^{-\delta^{+}_{12}},~~~~~~~~~~\\
{\cal I}^{-}_{\bf Bulk}(12,h,c)&=&g~\pi^h~\Gamma\left(\frac{\Delta^{-}_1+\Delta^{-}_2+c-h}{2}\right)~\int^{\infty}_{0}~\frac{ds}{s}~~s^{{\displaystyle h+c-\sum_{(ij)}}^{'}\displaystyle \delta^{-}_{ij}}~(1+s^2)^{-\delta^{-}_{12}},~~~~~~~~~~\\
{\cal I}^{+}_{\bf Bulk}(34,h,-c)&=&g~\pi^h~\Gamma\left(\frac{\Delta^{+}_3+\Delta^{+}_4-c-h}{2}\right)~\int^{\infty}_{0}~\frac{d\bar{s}}{\bar{s}}~~\bar{s}^{{\displaystyle h-c-\sum_{(ij)}}^{'}\displaystyle \delta^{+}_{ij}}~(1+\bar{s}^2)^{-\delta^{+}_{34}},\\
{\cal I}^{-}_{\bf Bulk}(34,h,-c)&=&g~\pi^h~\Gamma\left(\frac{\Delta^{-}_3+\Delta^{-}_4-c-h}{2}\right)~\int^{\infty}_{0}~\frac{d\bar{s}}{\bar{s}}~~\bar{s}^{{\displaystyle h-c-\sum_{(ij)}}^{'}\displaystyle \delta^{-}_{ij}}~(1+\bar{s}^2)^{-\delta^{-}_{34}}.\eea
Further,  one can simplify the expressions for the Mellin-Barnes amplitude integrals in terms of the Mandelstam variables in de Sitter space as:
\bea {\cal M}^{+}_4(s_{12})=\frac{g^2}{
\displaystyle \Gamma\left(\frac{\Delta^{+}_1+\Delta^{+}_2-s_{12}}{2}\right)\Gamma\left(\frac{\Delta^{+}_3+\Delta^{+}_4-s_{12}}{2}\right)}~ \int^{+i\infty}_{-i\infty}\frac{dc}{2\pi i}~\frac{{\cal R}^{+}_h(c){\cal R}^{+}_h(-c)}{\left\{\left(\delta^{+}-h\right)^2-c^2\right\}},~~~~~~\eea\bea
{\cal M}^{-}_4(s_{12})=\frac{g^2}{
\displaystyle \Gamma\left(\frac{\Delta^{-}_1+\Delta^{-}_2-s_{12}}{2}\right)\Gamma\left(\frac{\Delta^{-}_3+\Delta^{-}_4-s_{12}}{2}\right)}~ \int^{+i\infty}_{-i\infty}\frac{dc}{2\pi i}~\frac{{\cal R}^{-}_h(c){\cal R}^{-}_h(-c)}{\left\{\left(\delta^{-}-h\right)^2-c^2\right\}},~~~~~~\eea
where we define the functions ${\cal R}^{+}_h(c)$,  ${\cal R}^{-}_h(c)$,  ${\cal R}^{+}_h(-c)$ and ${\cal R}^{-}_h(-c)$ by the following expressions:
\bea {\cal R}^{+}_h(c):&=&\frac{1}{2\Gamma(c)}~\Gamma\left(\frac{\Delta^{+}_1+\Delta^{+}_2+c-h}{2}\right)\Gamma\left(\frac{\Delta^{+}_3+\Delta^{+}_4+c-h}{2}\right),\\
{\cal R}^{-}_h(c):&=&\frac{1}{2\Gamma(c)}~\Gamma\left(\frac{\Delta^{-}_1+\Delta^{-}_2+c-h}{2}\right)\Gamma\left(\frac{\Delta^{-}_3+\Delta^{-}_4+c-h}{2}\right),\eea
and 
\bea 
{\cal R}^{+}_h(-c):&=&\frac{1}{2\Gamma(-c)}~\Gamma\left(\frac{\Delta^{+}_1+\Delta^{+}_2-c-h}{2}\right)\Gamma\left(\frac{\Delta^{+}_3+\Delta^{+}_4-c-h}{2}\right),\\
{\cal R}^{-}_h(-c):&=&\frac{1}{2\Gamma(-c)}~\Gamma\left(\frac{\Delta^{-}_1+\Delta^{-}_2-c-h}{2}\right)\Gamma\left(\frac{\Delta^{-}_3+\Delta^{-}_4-c-h}{2}\right).\eea
Finally,  the above mentioned  Mellin-Barnes amplitude integrals can be expressed after performing the integral in complex plane as:
\bea {\cal M}^{+}_4(s_{12})&=&\frac{g^2}{2}~\frac{1}{(s_{12}-\delta^{+})}~\frac{1}{\Gamma(1+\delta^{+}-h)}~\Gamma\left(\frac{\Delta^{+}_1+\Delta^{+}_2+\delta^{+}-h}{2}\right)\Gamma\left(\frac{\Delta^{+}_3+\Delta^{+}_4+\delta^{+}-h}{2}\right)\nonumber\\
&&\times {}_3F_1\left(\frac{2+\delta^{+}-\Delta^{+}_1-\Delta^{+}_2}{2},\frac{2+\delta^{+}-\Delta^{+}_3-\Delta^{+}_4}{2},\frac{\delta^{+}-s_{12}}{2},\frac{2+\delta^{+}-s_{12}}{2},1+\delta^{+}-h,1\right).\nonumber\\
&=&\sum^{\infty}_{n=0}\frac{P^{\delta^{+}}_n}{s_{12}-\delta^{+}-2n}{\cal V}^{\Delta^{+}_1,\Delta^{+}_2,\delta^{+}}_{[0,0,n]}{\cal V}^{\Delta^{+}_3,\Delta^{+}_4,\delta^{+}}_{[0,0,n]},\\
{\cal M}^{-}_4(s_{12})&=&\frac{g^2}{2}~\frac{1}{(s_{12}-\delta^{-})}~\frac{1}{\Gamma(1+\delta^{-}-h)}~\Gamma\left(\frac{\Delta^{-}_1+\Delta^{-}_2+\delta^{-}-h}{2}\right)\Gamma\left(\frac{\Delta^{-}_3+\Delta^{-}_4+\delta^{-}-h}{2}\right)\nonumber\\
&&\times {}_3F_1\left(\frac{2+\delta^{-}-\Delta^{-}_1-\Delta^{-}_2}{2},\frac{2+\delta^{-}-\Delta^{-}_3-\Delta^{-}_4}{2},\frac{\delta^{-}-s_{12}}{2},\frac{2+\delta^{-}-s_{12}}{2},1+\delta^{-}-h,1\right).\nonumber\\
&=&\sum^{\infty}_{n=0}\frac{P^{\delta^{-}}_n}{s_{12}-\delta^{-}-2n}{\cal V}^{\Delta^{-}_1,\Delta^{-}_2,\delta^{-}}_{[0,0,n]}{\cal V}^{\Delta^{-}_3,\Delta^{-}_4,\delta^{-}}_{[0,0,n]},\eea 
where the three point vertices and the normalised propagators for two branches of solutions in de Sitter space are defined as:
\bea {\cal V}^{\Delta^{+}_1,\Delta^{+}_2,\delta^{+}}_{[0,0,0]}&=&g~\Gamma\left(\frac{\Delta^{+}_1+\Delta^{+}_2+\delta^{+}-2h}{2}\right),\\
{\cal V}^{\Delta^{+}_3,\Delta^{+}_4,\delta^{+}}_{[0,0,0]}&=&g~\Gamma\left(\frac{\Delta^{+}_3+\Delta^{+}_4+\delta^{+}-2h}{2}\right),\eea\bea
{\cal V}^{\Delta^{+}_1,\Delta^{+}_2,\delta^{+}}_{[0,0,n]}&=&{\cal V}^{\Delta^{+}_1,\Delta^{+}_2,\delta^{+}}_{[0,0,0]}~\frac{\displaystyle \Gamma\left(1-\frac{1}{2}\left(\Delta^{+}_1+\Delta^{+}_2-\delta^{+}\right)+n\right)}{\displaystyle \Gamma\left(1-\frac{1}{2}\left(\Delta^{+}_1+\Delta^{+}_2-\delta^{+}\right)\right)}\nonumber\\
&=&g~\Gamma\left(\frac{\Delta^{+}_1+\Delta^{+}_2+\delta^{+}-2h}{2}\right)~\frac{\displaystyle \Gamma\left(1-\frac{1}{2}\left(\Delta^{+}_1+\Delta^{+}_2-\delta^{+}\right)+n\right)}{\displaystyle \Gamma\left(1-\frac{1}{2}\left(\Delta^{+}_1+\Delta^{+}_2-\delta^{+}\right)\right)},\\
{\cal V}^{\Delta^{+}_3,\Delta^{+}_4,\delta^{+}}_{[0,0,n]}&=&{\cal V}^{\Delta^{+}_3,\Delta^{+}_4,\delta^{+}}_{[0,0,0]}~\frac{\displaystyle \Gamma\left(1-\frac{1}{2}\left(\Delta^{+}_3+\Delta^{+}_4-\delta^{+}\right)+n\right)}{\displaystyle \Gamma\left(1-\frac{1}{2}\left(\Delta^{+}_3+\Delta^{+}_4-\delta^{+}\right)\right)}\nonumber\\
&=&g~\Gamma\left(\frac{\Delta^{+}_3+\Delta^{+}_4+\delta^{+}-2h}{2}\right)~\frac{\displaystyle \Gamma\left(1-\frac{1}{2}\left(\Delta^{+}_3+\Delta^{+}_4-\delta^{+}\right)+n\right)}{\displaystyle \Gamma\left(1-\frac{1}{2}\left(\Delta^{+}_3+\Delta^{+}_4-\delta^{+}\right)\right)},
\eea 
and
\bea {\cal V}^{\Delta^{-}_1,\Delta^{-}_2,\delta^{-}}_{[0,0,0]}&=&g~\Gamma\left(\frac{\Delta^{-}_1+\Delta^{-}_2+\delta^{-}-2h}{2}\right),\\
{\cal V}^{\Delta^{-}_3,\Delta^{-}_4,\delta^{-}}_{[0,0,0]}&=&g~\Gamma\left(\frac{\Delta^{-}_3+\Delta^{-}_4+\delta^{-}-2h}{2}\right),\\
{\cal V}^{\Delta^{-}_1,\Delta^{-}_2,\delta^{-}}_{[0,0,n]}&=&{\cal V}^{\Delta^{-}_1,\Delta^{-}_2,\delta^{-}}_{[0,0,0]}~\frac{\displaystyle \Gamma\left(1-\frac{1}{2}\left(\Delta^{-}_1+\Delta^{-}_2-\delta^{-}\right)+n\right)}{\displaystyle \Gamma\left(1-\frac{1}{2}\left(\Delta^{-}_1+\Delta^{-}_2-\delta^{-}\right)\right)}\nonumber\\
&=&g~\Gamma\left(\frac{\Delta^{-}_1+\Delta^{-}_2+\delta^{-}-2h}{2}\right)~\frac{\displaystyle \Gamma\left(1-\frac{1}{2}\left(\Delta^{+}_1+\Delta^{-}_2-\delta^{-}\right)+n\right)}{\displaystyle \Gamma\left(1-\frac{1}{2}\left(\Delta^{-}_1+\Delta^{-}_2-\delta^{-}\right)\right)},\\
{\cal V}^{\Delta^{-}_3,\Delta^{-}_4,\delta^{-}}_{[0,0,n]}&=&{\cal V}^{\Delta^{-}_3,\Delta^{-}_4,\delta^{-}}_{[0,0,0]}~\frac{\displaystyle \Gamma\left(1-\frac{1}{2}\left(\Delta^{-}_3+\Delta^{-}_4-\delta^{-}\right)+n\right)}{\displaystyle \Gamma\left(1-\frac{1}{2}\left(\Delta^{-}_3+\Delta^{-}_4-\delta^{-}\right)\right)}\nonumber\\
&=&g~\Gamma\left(\frac{\Delta^{-}_3+\Delta^{-}_4+\delta^{-}-2h}{2}\right)~\frac{\displaystyle \Gamma\left(1-\frac{1}{2}\left(\Delta^{-}_3+\Delta^{-}_4-\delta^{-}\right)+n\right)}{\displaystyle \Gamma\left(1-\frac{1}{2}\left(\Delta^{-}_3+\Delta^{-}_4-\delta^{-}\right)\right)},
\eea 
Also the normalization factors $P^{\Delta^{+}}_n$ and $P^{\Delta^{-}}_n$ are defined as:
\bea P^{\Delta^{+}}_n&=& \frac{1}{2n! ~\Gamma(1+\delta^{+}-h+n)}~~~~~~~~~{\rm where}~~~h=\frac{d}{2},~\delta^{+}=\frac{\Delta^{+}}{2},~~~~\\
P^{\Delta^{-}}_n&=& \frac{1}{2n! ~\Gamma(1+\delta^{-}-h+n)}~~~~~~~~~{\rm where}~~~h=\frac{d}{2},~\delta^{-}=\frac{\Delta^{-}}{2}.~~~~\eea
Here at the end the Mellin amplitude in de Sitter space can be expressed as an infinite sum of products of three point vertices and propagators individually coming from two branches, $+$ and $-$ of solutions. 

\section{Summary and Conclusion}
\label{sec:conclu}

The key findings of this paper are appended below point-wise:
\begin{itemize}
    \item The study investigated CFT correlation functions derived in the AdS/CFT and dS/CFT settings using the Mellin formalism, with promising results. 

    \item In contrast to the complicated D-functions that develop in coordinate space, contact interactions have simple polynomial Mellin amplitudes. Even the dreaded stress-tensor exchange diagram is reduced to a simple rational function for sparsely coupled scalars. In the Mellin representation, explicit gamma functions capture double-trace operators corresponding to external leg fusion, but single-trace operators and their progeny corresponding to internal lines or bulk-to-bulk propagators appear as simple poles of the Mellin Barnes amplitude.

    \item These fundamental analytic features of Mellin amplitudes also indicate which operators propagate across a particular Witten diagram in AdS space or a Witten-like diagram in dS space. We used the Mellin framework to construct tree-level correlation functions of generic scalars in the (d+1)-dimensional de Sitter space. This covers both n-point contact and four-point exchange diagrams.

    \item From an observational standpoint, computations conducted, particularly from the dS/CFT perspective, are immensely important in the context of studying primordial cosmic correlations. Using the Mellin-Barnes model of quasi-dS correlation functions with $d=3$, one may investigate numerous hitherto undiscovered aspects of tiny and large primordial fluctuations, which are closely relevant to the study of the inflationary paradigm and primordial black hole generation. 

    \item  Though we did not explicitly compute such higher-point cosmic correlation functions in this study, our findings for the dS/CFT correlation functions might be expanded to investigate other unresolved topics in the current context of discussion. There are a couple of other directions in which the results obtained for dS/CFT correlators can be extended, such as studying cosmological collider signals in terms of non-Gaussian cosmological correlation functions and, last but not least, non-perturbative treatment of bootstrapping cosmological correlators.

    \item We have explicitly computed the expression for the three-point function and the associated amplitudes using the Mellin-Barnes representation in $d+1$-dimensional dS and AdS space-time. 

    \item Finally,  we have computed the four-point function and the associated amplitudes using the Mellin-Barnes representation in $d+1$-dimensional dS and AdS space-time. One can, in principle, compute the higher point correlators and the associated Mellin-Barnes amplitudes in AdS and dS space. However, due to having the tight constraints from cosmological observations, finding out the expressions for the higher-point cosmological correlators, more than the four-point, is not physically relevant. Since the results obtained for dS Mellin Barnes amplitudes are cosmologically relevant, in the present context of discussion, we have restricted our computation to four-point function. 
\end{itemize}
In this paper, we have restricted our analysis for scalar fields only. In the future version of this work, we have a plan to extend the present analysis for any arbitrary spin-$s$ particle. Further, the Mellin formalism's ability to reveal analytic characteristics of late-time correlators at the tree level drives the study of quantum corrections in this context. A bootstrap approach to de Sitter correlators should help better comprehend quantum corrections at both the perturbative and non-perturbative levels. After exploring four-point correlators in AdS and dS, it would be fascinating to investigate the more complex issue of flat space holography within this framework. It would be intriguing to learn more about the potential relationship between the findings in this research and the related, intermediate flat space analysis. To understand more about this issue please consider refs. \cite{Pasterski:2016qvg,Pasterski:2017kqt,Cardona:2017keg}.


\acknowledgments

 SC would like to thank The North American Nanohertz Observatory for Gravitational
Waves (NANOGrav) collaboration and the National Academy of Sciences (NASI), Prayagraj, India, for being elected
as an associate member and the member of the academy respectively. SC would also like to thank all the members
of Quantum Aspects of the Space-Time \& Matter (QASTM) for elaborative discussions. Last but not least, we
acknowledge our debt to the people belonging to the various parts of the world for their generous and steady support
for research in natural sciences.

\newpage
\addcontentsline{toc}{section}{References}
\bibliographystyle{utphys}
\bibliography{references}

\end{document}